\documentclass[%
 reprint,
 superscriptaddress,
 amsmath,amssymb,
aps,
]{revtex4-2}

\usepackage{graphicx}
\usepackage{dcolumn}
\usepackage{bm}
\usepackage[colorlinks=true,linkcolor=blue,citecolor=blue,urlcolor=blue]{hyperref}

\usepackage{float}
\usepackage{xcolor}
\usepackage{amsmath} 
\usepackage{mathtools}
\usepackage{caption}
\usepackage[utf8]{inputenc}
\usepackage{ragged2e}
\usepackage[english]{babel}
\usepackage{multirow}
\usepackage[subrefformat=parens,labelformat=parens]{subfig}
\usepackage[capitalise,noabbrev,nameinlink]{cleveref}

\begin{document}

\preprint{APS/123-QED}

\title{Chip-integrated Spectroscopy Capable of Temperature Retrieval}

\author{Yifan Du}
\affiliation{%
Department of Physics, Stevens Institute of Technology,\\
 1 Castle Point Terrace, Hoboken, New Jersey, 07030, USA
}
\affiliation{%
 Center for Quantum Science and Engineering, Stevens Institute of Technology,\\
1 Castle Point Terrace, Hoboken, New Jersey 07030, USA }

\author{Yong Meng Sua}
\affiliation{%
Department of Physics, Stevens Institute of Technology,\\
 1 Castle Point Terrace, Hoboken, New Jersey, 07030, USA
}
\affiliation{%
 Center for Quantum Science and Engineering, Stevens Institute of Technology,\\
1 Castle Point Terrace, Hoboken, New Jersey 07030, USA }
\affiliation{Quantum Computing Inc., 5 Marinview Plaza, Hoboken, New Jersey 07030, USA}

\author{Santosh Kumar}
\affiliation{%
Department of Physics, Stevens Institute of Technology,\\
 1 Castle Point Terrace, Hoboken, New Jersey, 07030, USA
}
\affiliation{%
 Center for Quantum Science and Engineering, Stevens Institute of Technology,\\
1 Castle Point Terrace, Hoboken, New Jersey 07030, USA }

\author{Jiuyi Zhang}
\affiliation{%
Department of Physics, Stevens Institute of Technology,\\
 1 Castle Point Terrace, Hoboken, New Jersey, 07030, USA
}
\affiliation{%
 Center for Quantum Science and Engineering, Stevens Institute of Technology,\\
1 Castle Point Terrace, Hoboken, New Jersey 07030, USA }

\author{Xiangzhi Li}
\affiliation{%
Department of Physics, Stevens Institute of Technology,\\
 1 Castle Point Terrace, Hoboken, New Jersey, 07030, USA
}
\affiliation{%
 Center for Quantum Science and Engineering, Stevens Institute of Technology,\\
1 Castle Point Terrace, Hoboken, New Jersey 07030, USA }

\author{Yongxiang Hu}
\affiliation{NASA Langley Research Center,
Hampton, Virginia 23681, USA}

\author{Parminder Ghuman}
\affiliation{NASA Earth Science Technology Office, Greenbelt, Maryland 20771, USA}

\author{Yuping Huang}
\email{yhuang5@stevens.edu}
\affiliation{%
Department of Physics, Stevens Institute of Technology,\\
 1 Castle Point Terrace, Hoboken, New Jersey, 07030, USA
}
\affiliation{%
 Center for Quantum Science and Engineering, Stevens Institute of Technology,\\
1 Castle Point Terrace, Hoboken, New Jersey 07030, USA }

\date{\today}
\begin{abstract}
We demonstrate a chip-integrated emission spectroscope capable of retrieving the temperature of the light sources. It consists of a single photon detector with low dark counts and a sweeping on-chip filter with 2 pm spectral resolution in the visible and near-infrared regimes. With wildfire sensing applications in mind, we test our system with a hollow cathode lamp to simulate the K-line emission, and show how the models of Doppler and collision broadening in the plasma can be used for temperature retrieval. With favorable device parameters, high spectral resolution, and a novel temperature retrieval capability, our technique may find broad applications in environmental monitoring, astrophysics, plasma physics, and so on.
\end{abstract}

\maketitle


\section{\label{sec:level1}INTRODUCTION}

Precise and reliable measurements of emission lines from light sources such as wildfire, celestial bodies, and fusion reactors serve important tools for environmental monitoring, astrophysics, plasma physics, and so on \cite{K_wildfire, weight, James_Webb, plasma_diagnostics,Mars}. For example, trace element emissions from biomass burning have been studied for wildfire detection, for which, potassium lines around at 769.9 nm and 766.5 nm (associated respectively with the D$_1$ and D$_2$ transitions  \cite{NIST_K,D_doublet_accurate}), are strong due to its low ionization energy (4.34 eV). Aided by high atmospheric transmissivity \cite{K_trans}, long-distance wildfire sensing can be developed, on both airborne and spaceborne platforms \cite{K_wildfire}. However, the spectral resolution of existing devices is low (over 1~nm), which limits the measurement signal to noise and hinders the deployment in daytime due to the high solar background \cite{K_wildfire,Alaska_K,wildfire_temp2}. 

Another important application area is with plasma physics, particularly in fusion reactor diagnostics, where the helium number density is a critical parameter for assessing the energy required to heat the plasma under specific confinement conditions and for predicting the fusion rate \cite{Modern_physics,Tokamak_confine}. Excessive helium density could impede the fusion burn, creating a significant diagnostic challenge. Therefore, real-time measurement of helium density in the plasma core is essential. To this end, the emission spectra of plasma can provide ion density through profile intensity and temperature through linewidth. Yet the existing resolution is limited to 0.2 nm, which gives low intensity accuracy \cite{Tokamak_line_meas}.


Besides the demand for spectral resolution, wide deployment of remote spectroscopy measurement also calls for the systems' favorable Size, Weight, and Power (SWaP) parameters. This is particularly emphasized by space-based missions, as putting a heavy and bulky spectrometer in a satellite incurs exceedingly high costs  \cite{weight}.


Both of the above resolution and SWaP challenges can be addressed by new spectroscopy techniques based on photonic integrated circuits \cite{spectrometer_heater,broadband_spectrometer_LN,phosphorus_spectrometer,compact_photodetector,silicon_spectrometer,on_chip_spectro,spectro_Boyd}. Among these, the thin-film lithium niobate (TFLN) platform stands out due to its exceptionally low propagation loss, wide optical transparency window (near-UV to mid-IR), and efficient optical tuning using electro-/thermal-optical modulations \cite{Metal_absorb2, LN_science, EOM,chen}. Of particular relevance to spectroscopy are ultra-high-Q micro-ring resonators (MRR) in TFLN, for which an intrinsic Q-factor of $2.9\times10^7$ has been demonstrated \cite{High_Q}.  Recently, we used thermally tunable MRR on lithium niobate on insulator (LNOI) for CO$_2$ \cite{CO2} and O$_2$ \cite{O2} absorption spectroscopy to measure their atmosphere concentration, with millimeter sizes and picometer resolution. 

Here, we explore LNOI circuits for emission spectroscopy, and demonstrate a novel capability in retrieving the temperature of the light sources. Our system consists of a single photon detector with low dark counts and a sweeping LNOI MRR filter with 2 pm spectral resolution in the visible and near-infrared regimes. With the wildfire sensing application in mind, we test our system with a hollow cathode lamp (HCL) to emulate the K-line emission, and show how the models of Doppler and collision broadening in the plasma can be used for temperature retrieval. With favorable SWaP parameters, high resolution, and temperature retrieval capablility, our technique can find broad applications in environmental monitoring, astrophysics, plasma physics, and so on \cite{white_dwarf,AGN_collision}.


\section{Device Design and Calibration}
\begin{figure}[h!]
\captionsetup[subfloat]{position=top,justification=raggedright,singlelinecheck=false} 
\centering
\subfloat[][\label{fig:ring}]{\includegraphics[width=1.01\linewidth]{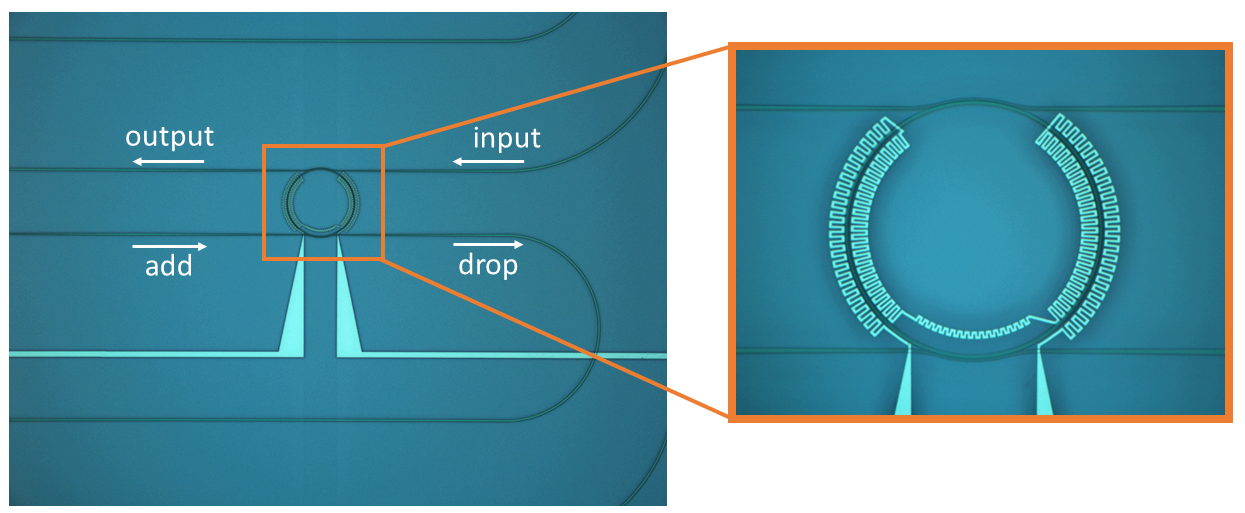}}\\
\subfloat[][\label{fig:scan}]{\includegraphics[width=\linewidth]{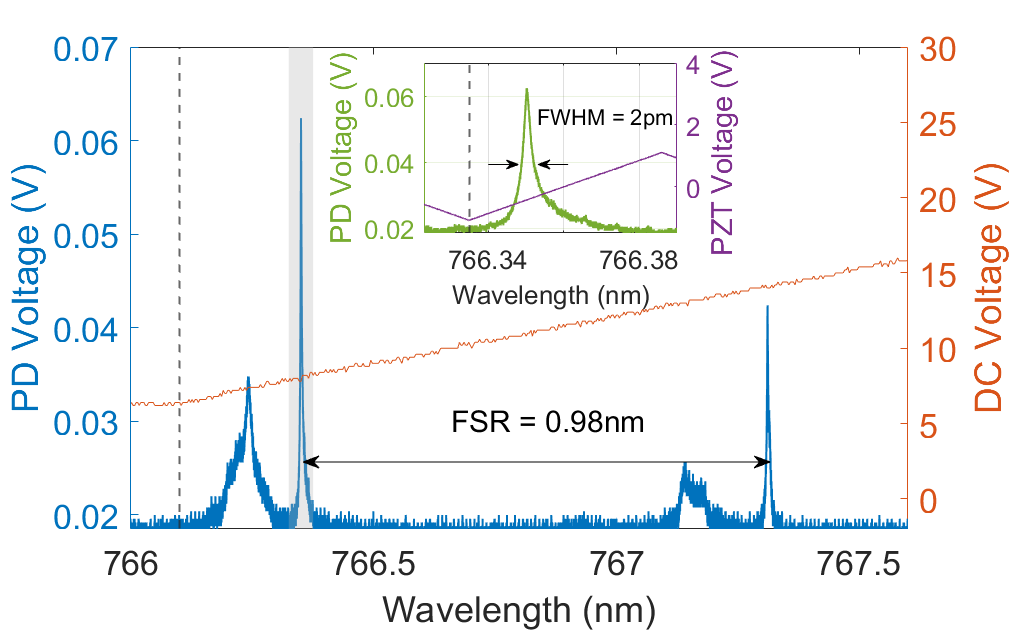}} 
\caption{\justifying (a) Microscope images of the MRR chip. (b) Calibraition results, where we show a coarse scan of laser output for resonance characterization, alongside a fine scan using external frequency modulation for linewidth characterization. The fine scan range (51 pm) is indicated as the grey shaded region, with the inset showing a 2-pm FWHM measurement. The dashed lines represent the starting point of the scan wavelength. The left y-axes of the figures represent the photodiode (PD) output voltage, which is linear proportional to the output optical power from the MRR drop port. The right y-axes of the main figure and the inset indicate the DC motor voltage and the piezoelectric transducer (PZT) voltage of the laser, respectively, for the calibration.}
\label{fig:laser scan}
\end{figure}

Figure~\subref*{fig:ring} shows the circuit layout of the MRR filter used in this demonstration. It consists of a high-Q ring resonator and two pulley bus waveguides on the two sides of the ring, respectively \cite{O2,CO2}. On one side are the input and output ports, and on the other sides are the add and drop ports. The incoming signal is injected into the input port, and the filtered signal is collected at the drop port. To sweep the filter center wavelength, a zigzag heater surrounding the ring is deposited on top of the MRR, so that its resonant wavelengths are thermally tuned. Here, the zigzag layout is to increase the heating area of the electrode and thus the heating efficiency. Alternatively, instead of the thermo-optic tuning, electro-optic tuning can be applied to utilize LN's large electro-optic coefficients. In comparison, the former is more power efficient, but the latter can achieve ultrahigh tuning speed (at hundreds of GHz, versus hundreds of KHz). 

Unlike typical interference filters, the MRR filter has periodic transmission spectral windows, separated by free-spectral range (FSR) given by \cite{silicon}
\begin{equation}\label{eq:FSR}
    \mathrm{FSR}=\frac{\lambda^2}{n_g2\pi R}
\end{equation}
where $\lambda$ is the center wavelength, $R$ is the ring radius, and $n_g$ is the group refractive index at the center wavelength. In typical settings, FSR is on the order of nanometers, so that the filter will admit not only the target signal photons, but also those in other periodic windows. Hence, if there is broad background noise, a coarse filter is needed to apply around the target wavelength for noise suppression.  

For the current experiment, we design and fabricate the circuits on a 300-nm TFLN wafer, shallow etched to a height of 180 nm. The ring radius is $R=40$ $\mathrm{\mu m}$. The pulley and ring waveguides have top widths of 280 nm and 480 nm, respectively, with a sidewall angle of 82 degree. The detailed fabrication process and the circuit dimensions of the MRR can be found in Appendix \ref{appendix:fab}. These waveguide dimensions ensure that only the fundamental TE mode propagates in the waveguides and all the other modes are suppressed. For the center wavelength $\lambda=770$ nm, the simulated group refractive index $n_g=2.4038$, so that the FSR is 0.98 nm by design. 

The microscope image of the fabricated chip is shown in Fig.~\subref*{fig:ring}. To characterize its transmission spectrum, we use a tunable laser (New Focus TLB-6700) whose output wavelength can be swept using both coarse and fine scans, the latter of which significantly enhances the calibration accuracy over our previous measurement to reduce the measurement error from 2 pm to 0.03 pm \cite{O2}. The details of the calibration method can be found in Appendix~\ref{appendix:cali}.

Because of the 0.98-nm FSR and the limited maximum fine scan range of 150 pm, we first use the laser internal coarse scan to locate the resonance wavelength. Then we perform a fine scan around the resonance wavelength with a scaning range of 51 pm. The measurement results are shown in Fig.~\subref*{fig:scan}, where two TE00 modes are shown with a 0.98 nm FSR, as simulated. The resonance profile has a FWHM of 2-pm as shown in the inset of Fig.~\subref*{fig:scan}, corresponding to a loaded cavity of $Q=0.38\times 10^{6}$. Also shown in the figure are two wide transmission peaks, which correspond to TM00 modes, which have much higher propagation losses due to the waveguide dimension. The wavemeter measurement details of MMR characterization can be found in Appendix~\ref{appendix:cali}.

In this experiment, we measure specifically the potassium emission lines for potential wildfire detection. To this end, we use a potassium hollow cathode lamp (HCL), which emits the K-line doublet at 766.5 nm and 769.9 nm, respectively. The spacing between the two is 3.4 nm, so that only one line can be transmitted by the MRR at a time when its resonance is thermo-optically tunned. To simultaneously detect both transmission lines, one may cascade two MRR's in serial, with each's resonance tuned to be at one line and the input port of a second MRR connected to the through port of the first. Because of the on-chip integration, there is no insertion loss between the two. By cascading more MRR's, signals in many emission lines can be measured simultaneously.  




In our chip design, the heating electrode will introduce additional loss to the cavity, due to the evanescent light absorption by the metal \cite{Metal_absorb1,Metal_absorb2}. Thus it is desirable to spatially separate the electrode and the waveguide, so that there is little evanescent light reaching the electrode. However, if the gap between the two is large, the heater efficiency will be weak, so that the wavelength tunability is reduced. To achieve both low loss (thus high spectral resolution) and high tunability, we use a zigzag-shaped heater to encircle the ring, to increase the heating area and thus its efficiency. Positioning the zigzag on both sides of the ring waveguide increases the heating area while reducing the temperature gradience to provide uniform heating.

\begin{figure}[h!]
         \centering \includegraphics[width=0.4\textwidth]{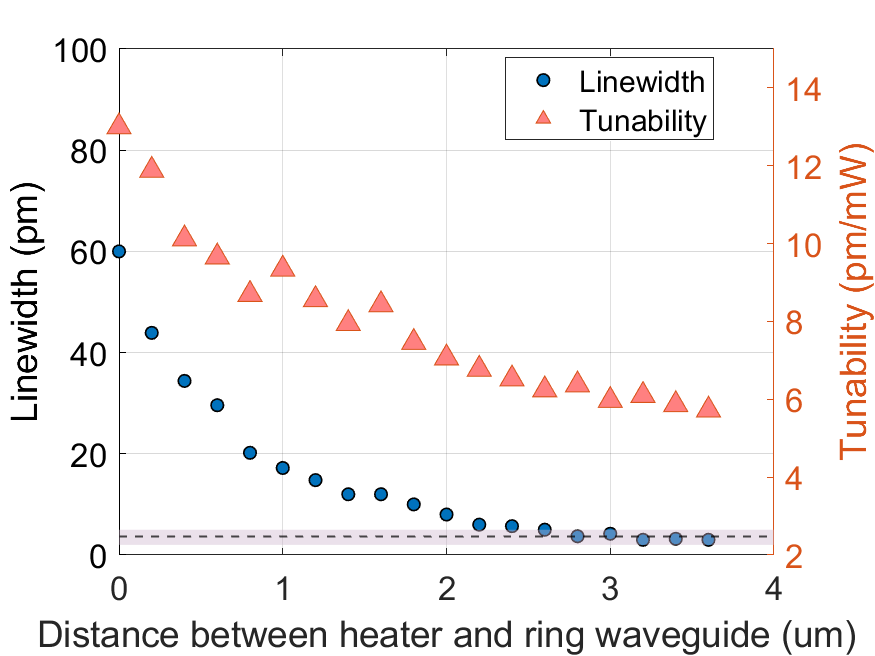}
    \caption{\justifying Resonance linewidth and tunability as functions of the lateral distance between the heater and waveguide. At the figure bottom, the dashed line and shaded area represent the mean linewidth and its standard deviation for the MRR without the heater deposited.}
        \label{fig:distance}
\end{figure}
\begin{figure}[h!]
         \centering \includegraphics[width=0.39\textwidth]{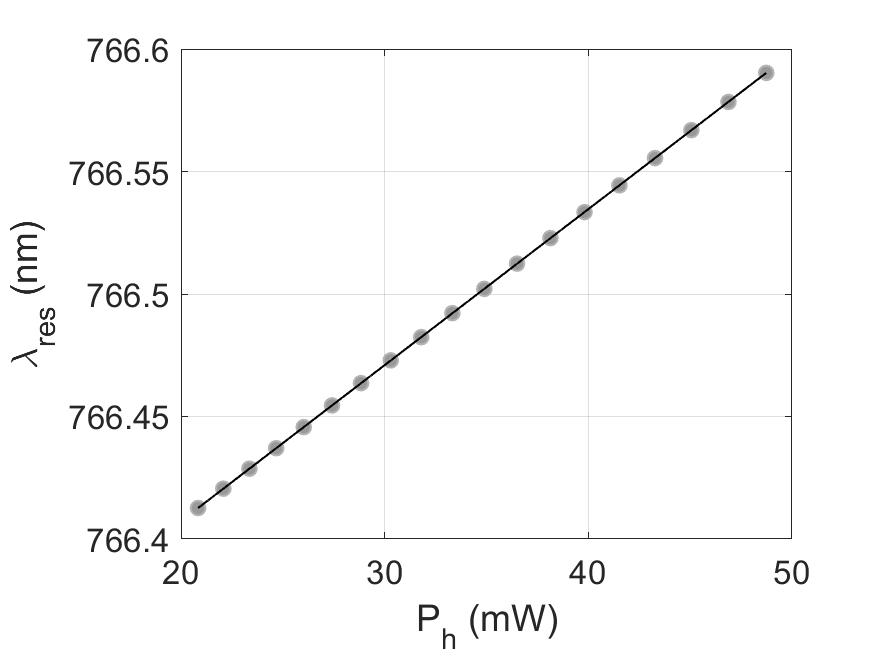}
    \caption{\justifying Resonance shift versus applied heating power. The scattered dots represent the measured data, while the solid line indicates the fitted result, with a slope of $6.3691$ pm/mW and coefficient of fitting determination $r^2=1$.}
        \label{fig:tunablity}
\end{figure}

Figure~\ref{fig:distance} shows the measured trade-off between the cavity linewidth and thermo-optic tunability when the lateral distance is varied between the electrode and waveguide. From the figure, when the distance exceeds 2.8 µm, the linewidth is nearly not affected by metal absorption. Yet the tunability remains sufficiently large for this experiment. In our current design, the distance is 2.8 µm as we are interested in measuring accurately the emission linewidth. To characterize the tunability, we apply voltage manually from 17 V to 26 V, at a step size of 0.5 V. For each voltage, the heater power is calculated using $P_{h}(V)=V^2/R_{h}$, where $R_{h}$ is the heater's resistance. The shifted resonance wavelength $\lambda_{res} (V)$ is measured using the tunable the laser. Figure~\ref{fig:tunablity} shows tunability of MRR with distance 2.8 $\mu m$ used for our experiment. The slope of the linear fitting line $\Delta\lambda_{res}/\Delta P_{h}$ indicates a tunability of $6.3691$ pm/mW. The coefficient of fitting determination, $r^2 = 1$, confirms the excellent linearity of the response. For other applications, one may choose an appropriate distance according to the trade-off, based on the specific application needs.  

\section{EXPERIMENT}

\begin{figure}[h!]
         \centering \includegraphics[width=0.5\textwidth]{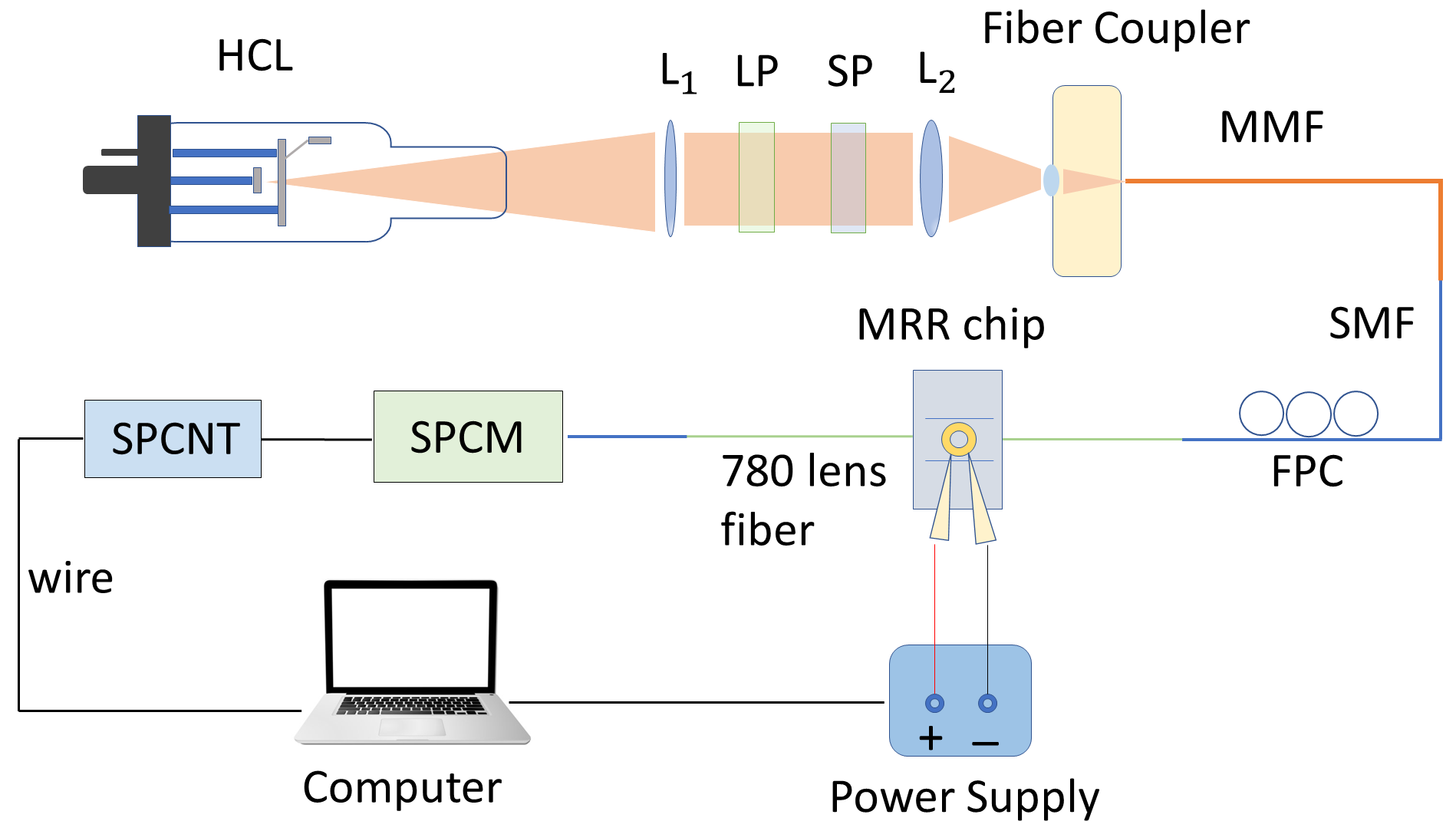}
    \caption{\justifying Experiment setup. HCL: hollow cathode lamp. L$_1$: thin lens with 100mm focal length. LP: long-pass filter. SP: short-pass filter. L$_2$: thin lens with 75mm focal length. MMF: multi-mode fiber (orange line). SMF: single-model fiber (blue line). FPC: fiber polarization controller. SPCM: single photon counting module. SPCNT: single photon counting device. The green lines denote the 780 nm lens fiber. The black lines represent electric wires.}
        \label{fig:setup1}
\end{figure}

The experiment for detecting potassium D lines is illustrated in Fig.~\ref{fig:setup1}. A hollow cathode lamp (HCL, Scinteck SI-HC-200042) is used as the light source for the K doublet emission. We apply a current of 8 mA and a voltage of 280 V to generate an intense electric field between the lamp's cathode and anode, where the inert Ne gas is ionized. The electrons emitted from the cathode, along with Ne ions, form a conducting plasma that excites potassium atoms to higher energy states. As the potassium atoms decay, they emit narrow D lines through a quartz window. A thin lens $L_1$ with a focal length of 100 mm collimates the diverging thermal light, which is then focused by a second lens $L_2$ with a 75 mm focal length. Short-pass (SP) and long-pass (LP) filters are employed to limit the spectral range to 750–800 nm, while effectively avoiding interference from other emission lines. The light is coupled into a multimode fiber (MMF) via a fiber coupler, where a single-mode fiber (SMF) is spliced with the MMF. The light passes through an fiber polarization controller (FPC) before coupled into the MRR chip via a lensed fiber. A power supply (GWINSTEK GPD-2303S) is used to sweep the resonance wavelength of the MRR by 200~pm by adjusting the applied voltage. When the resonance wavelength aligns with the potassium D-line, the emitted photons will pass through the MRR and reach a single-photon detector (PerkinElmer SPCM-AQR-14-FC-11260) through the drop port. The detector ouput pulses are registered by a counter (Thorlabs SPCNT) over a 1-second integration time to suppress Poissonian noise. During the process, a computer tunes the MRR resosnance wavelength by controlling the power supply and records the photon counts at the same time.

\begin{figure}[h!]
   \centering
\includegraphics[width=0.35\textwidth]{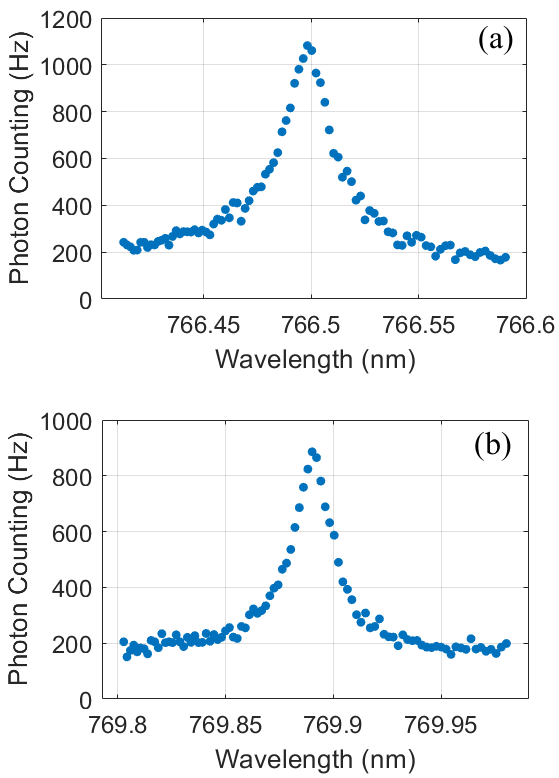}
   \caption{\justifying Single photon measurement for potassium (a) D2 line (b) D1 line. The filtered wavelengths of K doublet are 766.5 nm for D2 and 769.9 nm for D1. The dark count is $147\pm11$ Hz. }        \label{fig:D lines}
\end{figure}

The photon counting results are shown in Fig.~\ref{fig:D lines}, where (a) and (b) are for the Postassium D2 and D1 lines, respectively. For the D2 line, the power supply voltage is increased from 17 V to 26 V, at a step size of 0.1 V scanned automatically by the computer. Before the experiment, the calibrated resonance wavelength at the initial voltage $\lambda_{res}(V_0)$ and the tunability $\Delta\lambda_{res}/\Delta P_{h}$ were programmed into the computer for single photon detection. The results of D line profiles give photon counting versus power supply voltage. The computer automatically converts the voltage to the corresponding resonance wavelength by the relationship 
\begin{equation}
 \lambda_{res} (V)=\lambda_{res}(V_0)+\bigl(P_{h}(V)-P_{h}(V_0)\bigr)\frac{\Delta\lambda_{res}}{\Delta P_{h}}.
 \end{equation}
Hence, the accuracy of the calibrated tunability $\Delta\lambda_{res}/\Delta P_{h}$ and initial wavelength $\lambda_{res}(V_0)$, determines the precision of the measured wavelengths of the D-lines. The filtered wavelengths match well with the NIST database values of 766.5 nm and 769.9 nm \cite{NIST_K}. The dark count is measured as $147\pm 11$ Hz.

\section{TEMPERATURE RETRIEVAL BY LINEWIDTH}
Thus far, the reported method for temperature retrieval of wildfire relies on broadband spectral radiance fitted by the Planck function \cite{wildfire_temp2}. It is susceptible to errors due to the background spectrum of reflected solar radiance background which is also broadband. On the other hand, temperature in plasma is typically determined by the Boltzmann plot of multiple emission lines, which is prone to significant retrieval errors due to intensity fluctuations \cite{AGN_collision}. Similarly, temperature retrieval of stars using luminosity in the Stefan-Boltzmann equation requires stellar radius or angular diameter data \cite{star_temp1}, with radius or distance uncertainty affecting accuracy. Spectral contamination further impacts luminosity accuracy.

Here, we introduce a novel method of temperature retrieval using the fine measurement of emission linewidth with the MRR filter. In this experiment, by varying the lamp current from 8 mA to 40 mA, the cathode temperature is raised, which, in turn, boosts the emitted current from the cathode. This leads to a higher plasma charge density and, consequently, an elevated plasmon temperature \cite{plasma_temp_HCL}. The increase in temperature enhances the thermal motion of potassium atoms, resulting in greater Doppler broadening of the linewidth. Therefore, the increase of linewidth is observed as we increase the lamp current.

\begin{figure}[h!]
         \centering \includegraphics[width=0.45\textwidth]{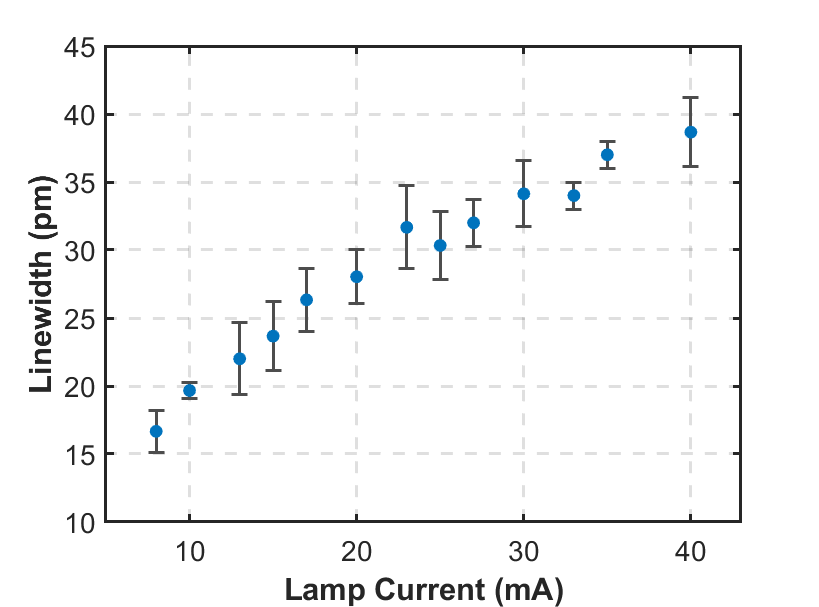}
    \caption{\justifying Measured linewidth vs lamp current. The lamp current is scanned from 8 mA to 40 mA. For each applied lamp current, the linewidth measurement was repeated three times. The scattered points represent the mean linewidth values, with error bars indicating the standard deviation across the repeated measurements.}
        \label{fig:linewidth}
\end{figure}

Figure~\ref{fig:linewidth} plots the change of linewidth as the lamp current is varied. The error bar represents the standard deviation of three repeated linewidth measurements taken at each lamp current. The observed linewidth is the result of a convolution between the Voigt profile of the emitted light from the lamp and the Lorentzian profile of the ring resonator. The Voigt profile is a convolution of Doppler broadening, caused by thermal motion, and collision broadening, which arises from the reduced lifetime of the excited state of potassium atoms due to collisions \cite{Modern_physics}. The total emission spectrum is therefore
 \begin{equation}
    V(\lambda,T) =G(\lambda,T) \otimes\bigl( L_1(\lambda,T)\otimes L_2(\lambda)\bigr)
    \label{eqn_V},
\end{equation}
where $G(\lambda, T)$ is the Gaussian profile describing the Doppler broadening, $L_1 (\lambda, T)$ is the lamp collision broadening Lorentzian profile, and $L_2 (\lambda)$ is the ring resonator Lorentzian profile. Associative property of convolution allows us to first convolve the lamp collision Lorentzian with the ring Lorentzian. This approach gives a convenience of the calculation, as the resulting combined Lorentzian profile has a linewidth just equal to the sum of the individual Lorentzian linewidths:
\begin{equation}\label{eq:lo_sum}
    \Delta\lambda_L=\Delta \lambda_1+\Delta \lambda_2
\end{equation}
where $\Delta\lambda_{1,2}$ are the FWHM of $L_{1,2}$, respectively. The collision broadening linewidth, as given by \cite{Loudon}, can be extended to address the specific conditions in this work. An illustration of the collision processes is shown in Fig.~\ref{fig:ionization} (a) in Appendix~\ref{appendix:insight}. Here, the Ne gas is partially ionized to form Ne ions. These Ne ions and electrons, present in equal densities, form a conducting plasma between the cathode and anode of the HCL.  
%
Therefore, we have $n=n_{Ne}+n_{Ne^+}$, where $n$ represents the total number density of particles, that consist of unionized neutral Ne with density $n_{Ne}$ and ionized Neon with density $n_{Ne^+}$. The density of Neon ions in plasma is equal to the density of electrons, expressed as $n_{Ne^+}=n_e$. The collision broadening linewidth is then modified to account for these plasma conditions as
\begin{equation}\label{eq:co}
    \begin{split}         \Delta\lambda_{1}&=\Delta\lambda_0+2\sqrt{2}\frac{\lambda^2}{c}n\sum_s\chi_s\sigma_{s-K} \Bar{v}_{s-K}
    \end{split}
\end{equation}
Here $\Delta\lambda_0$ is the potassium natural linewidth, which is $0.012$ pm \cite{K_natural}. The subscript $K$ denotes the potassium atoms, and $s$ represents the species that the potassium atoms collide with in the medium of the HCL glass tube, such as electrons (e), Ne ions (Ne$^+$), or neutral Ne atoms (Ne). The mole fraction of the species $s$ is denoted by $\chi_s$. The scattering cross section is given by $\sigma_{s-K}=\pi(a_s+a_K)^2$ where $a_s$ and $a_K$ are the radii of the mole fraction of the species $s$ and potassium atoms, respectively. $\Bar{v}_{s-K}$ is the mean relative speed, with  $\Bar{v}_{s-K}=\sqrt{\Bar{v}_s^2+\Bar{v}_K^2}$, where $\Bar{v}_s=\sqrt{8k_BT/\pi M_s}$, and $\Bar{v}_K=\sqrt{8k_BT/\pi M_K}$. $M_s$ and $M_K$ are masses of species s and potassium atoms, respectively. $k_B$ is the Boltzmann constant. In this context, the relations $\chi_{Ne}+\chi_e=1$ and $\chi_e=\chi_{Ne+}$ hold. Given that the lamp plasma is considered to be weakly ionized and assuming local thermal equilibrium for particles in the glass tube, the fraction of Ne ions $\chi_{Ne+}$, or the electron fraction $\chi_e$, also referred to as the degree of ionization, can be related to the temperature via the Saha equation \cite{Saha}
\begin{equation}
    \frac{\chi_e^2}{1-\chi_e}=\frac{2}{n\lambda_{\text{th}}^3}\frac{g_1}{g_0}\text{exp}\biggl(\frac{-\epsilon}{k_B T}\biggr)
\end{equation}
in which the thermal de Broglie wavelength is $\lambda_{\text{th}}=h/\sqrt{2\pi M_e k_B T}$. The degeneracy ratio of Neon ion to neutral Neon is given by $g_1/g_0=4$ \cite{Ne}. 
The first ionization energy of Ne is $\epsilon=21.56454$ eV \cite{Ne_ionization}.  In the retrieval calculation we use neutral Ne and Ne ion radii $a_{Ne}=0.02$ nm and $a_{Ne+}=0.112$ nm, Ne mass $M_{Ne}=M_{Ne+}=20.1797$ amu, potassium atomic radius  $a_{K}=0.235$ nm, potassium mass $M_{K}=39.0983$ amu, electron radius  $a_{e}=2.8179\times 10^{-6}$ nm, and electron mass $M_{e}=9.11\times 10^{-31}$ kg. Because the sealed glass tube has a constant volume and inert gas Ne particle numbers, we can determine the total number density, which is independent of temperature, by applying the ideal gas law, $n=p/kT$. By using $p=5$ torr at $T=300$ K, the calculated total number density is $n=1.61\times 10^{23}$ m$^{-3}$, which is on same order as reported in Ref. \cite{AGN_collision}. 

The Doppler broadening linewidth is defined as \cite{Loudon} 
\begin{equation}\label{eq:Dop}
    \Delta\lambda_G(T)=\frac{2\lambda}{c}\sqrt{ \frac{2\ln (2)k_BT}{M_K}}
\end{equation}
Then, by using the approximation relation between the FWHM of Voigt, Gaussian, and Lorentzian profiles, the FWHM of Voigt spectrum in Eq.~\labelcref{eqn_V} is then given to a very good approximation as \cite{Olivero}:
\begin{equation}\label{eq:Olivero}
\Delta\lambda_V=0.5346\Delta\lambda_L+\sqrt{0.2166\Delta\lambda_L^2+\Delta\lambda_G^2}.
\end{equation}

\begin{figure}[h!]
   \centering
\includegraphics[width=0.45\textwidth]{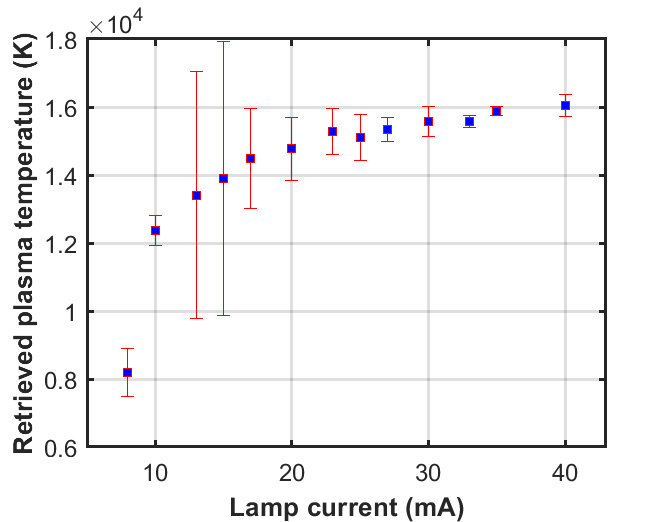}
   \caption{\justifying Retrieved temperature vs the lamp current with the plasma model incorporated, where the error bar is calculated through error propagation using the data in Fig.~\ref{fig:linewidth}.}   \label{fig:temp_plasma}
\end{figure}

The retrieved temperature using plasma model is shown in Fig.~\ref{fig:temp_plasma}.
As a comparison, the retrieved temperature is on the same order of magnitude ($\times 10^4$ K) of the temperature retrieved by Boltzmann plot in Ref.~\cite{AGN_collision}, where a similar hollow cathode discharge was employed for a laboratory-scale astrophysics experiment. To verify the validity of the retrieval model and gain insight into broadening mechanisms, we analyze the plasma ionization and the broadening competition in Appendix~\ref{appendix:insight}. Note that this retrieved temperature is almost one order of magnitude less than the model without considering plasma (completely unionized); see more discussions in Appendix~\ref{appendix:no plasma HCL}. 

\section{APPLICATIONS}
The demonstrated single-photon counting MRR spectroscopy have broad applications, ranging from detecting wildfires, to measuring fusion plasma and celestial bodies such as stars and black hole accretion disks in active galactic nuclei (AGN). For those applications, distinct circuit designs and retrieval models may be required to achieve the optimal performance. Table~\ref{table:compare} lists the emission bands, temperature range, emission linewidth, underlying broadening mechanisms, and some key physical properties of various light sources. As seen, most of the bands fall entirely or partially in the transparency window of lithium niobate, which spans from the UV to the mid-IR (MIR) (0.4-5 $\mu m$) \cite{Metal_absorb2, LN_science}. This provides an opportunity of applying the current LNOI MRR circuits to those various uses. For all of them, the 2~pm FWHM spectral resolution of the current MRR devices is adequate to quantitatively measure the line broadening effects, from which the source temperature can be retrieved. 
In Appendix~\ref{appendix:app}, we discuss several simple models for the temperature retrieval in those applications.

\begin{table*}
\caption{\label{tab:table3}Broadening linewidth comparison for different sources.}
\label{table:compare}
\begin{ruledtabular}
\begin{tabular}{m{1.5cm} m{2.7cm} m{2.8cm} m{2.2cm} m{1.3cm} m{1.3cm} m{1.8cm} m{1.5cm}} 
 Source & Emission bands & Temperature (Kelvin) & Pressure/Gas number density  & Doppler broadening (pm) & Collision broadening (pm) & Other broadening (pm)& Composited linewidth (pm)\\  
 \hline
 Potassium HCL & 766.5, 769.9 nm & $10^2$-$10^4$ & 1-10 torr/$10^{22}$-$10^{23}$ m$^{-3}$ & 2-25 & 0.25-2 & NA & 2-26\\  
 \hline
 Wildfire & 766.5, 769.9 nm (K) & $1000$-$2500$ \cite{K_trans,wildfire_temp2,wildfire_temp1} & 760 torr & 3-5 & 9-16 & NA & 11-17 \\
 &  423 nm (Ca) \cite{K_wildfire} &  &  &  &  &  &  \\
  & 589.3 nm (Na) \cite{K_wildfire} &  &  &  &  & & \\
\hline
 Fusion reactor & 518.9-533.1 nm \cite{Tokamak_line_meas} & $10^7$-$10^8$ \cite{Tokamak_confine,Overview_Tokamak} & $10^{19}$-$10^{21}$ m$^{-3}$ \cite{Overview_Tokamak,Greenwald_density,Plasma_spectroscopy} & $10^2$ & $<1$ & $10^{-4}$-$10^2$\cite{Stark_Tokyo,Stark_Griem}\footnotemark[1] & NA\\
 \hline
 Star & 420-690 nm &$ 10^3$-$10^5$ \cite{star_temp1,Star_temp2}& $10^{18}$-$10^{26}$ m$^{-3}$\cite{star_density1,star_density2} & 5-50 & $10^{-1}$-$10^2$ \cite{star_density2} & 0-$10^2$ \cite{Rotational_linewidth}\footnotemark[2] & NA  \\
 \hline 
AGN & UV-MIR (accretion disk) \cite{AGN_bands}& $10^5$-$10^9$ \cite{AGN_temp1} & $10^{15}$-$10^{23}$ m$^{-3}$ \cite{AGN_turbo} & $10^1$-$10^3$ \cite{AGN_width} & $10^{-1}$-$10^2$ \cite{AGN_collision} & $10^3$-$10^4$ \cite{AGN_width}\footnotemark[2]& NA\\
\end{tabular}
\end{ruledtabular}
\footnotetext[1]{Primarily attributed to Stark broadening \cite{Stark_Griem}.}
\footnotetext[2]{Primarily attributed to Stark broadening \cite{AGN_collision}, rotational broadening \cite{Rotational_linewidth}, and turbulence broadening \cite{AGN_turbo}.}
\end{table*}



Furthermore, different applications have different measurement requirements not only on wavelength bands and resolution, but also the scanning range and speed. For the thermo-optic tuning, the tunability  is given by \cite{heater_Lithium}
\begin{equation}
    \frac{\Delta\lambda_{res}}{\Delta P_{h}}=\frac{\lambda_{res}L_{h}}{n_{eff}2\pi R}\cdot\biggl(\frac{\partial n_{eff}}{\partial T}\biggr)\cdot\biggl(\frac{\Delta T}{\Delta P_{h}}\biggr).
    \label{eq_tunning}
\end{equation}
Here, $\Delta P_h$ is the change of the applied electric power for the heater and $\Delta\lambda_{res}$ is the resultant shift in the resonant wavelength. $L_{h}$ is the heater length. $R$ is the ring radius. $\Delta T$ is the temperature change of the heated ring waveguide. $\partial n_{eff}/\partial T$ is the thermal-optic coefficient.  $\Delta T/\Delta P_{h}$ is the heater efficiency (i.e., thermal resistance).

\begin{figure}[h!]
         \centering \includegraphics[width=0.41\textwidth]{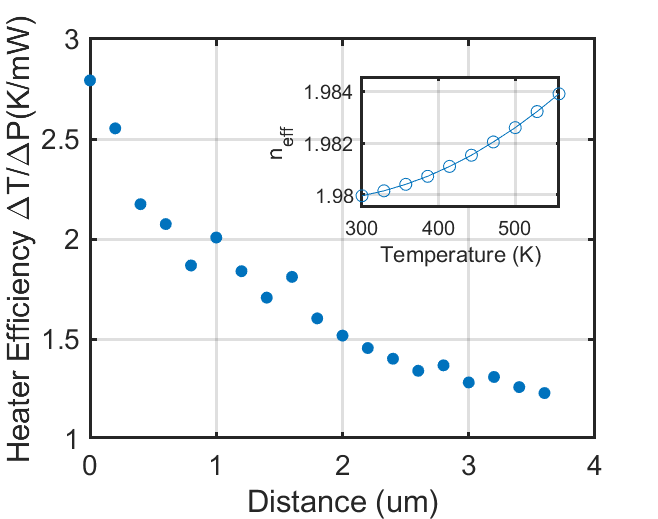}
    \caption{\justifying Heater efficiency vs distance for 770 MRR. The simulated thermal-optics coefficient is also shown as the inset. The temperature range of the heated ring waveguide ($300-557$ K) is obtained by using the applied voltage range of 0-32 V and the extracted heater efficiency.}
        \label{fig:heater efficiency}
\end{figure}

In Eq.~(\ref{eq_tunning}), the thermal-optic coefficient can be calculated by computing the relevant transverse mode properties using finite-difference time-domain (FDTD) simultations. In our case, $\partial n_{eff}/\partial T=1.79\times10^{-5}$ K$^{-1}$. The heater efficiency can can be efficiently adjusted by changing the distance between the heater and the MRR waveguide. Figure~\ref{fig:heater efficiency} and its inset plot the heater efficiency versus distance derived from Fig.~\ref{fig:distance} and Eq.~(\ref{eq_tunning}), and $n_{eff}$ as a function of the temperature from the FDTD simulation.    

The heater efficiency depends on the distance and thermal conductivity of the cladding material between the heater and waveguide, and the heater's electrode material composition and geometry \cite{silicon,spectrometer_heater}. It is, however, nearly independent of the waveguide geometry, suggesting that the above result can serve as a good design guideline empirically across a broad range of wavelengths. Finally, despite the perodicity in the MRR's resonant lines, the spectral measurement range can exceed the FSR by adding a tunable coarse filter with bandwidth narrower than FSR to block cross-talk interference from unwanted channels.



\section{CONCLUSION}
In conclusion, we have demonstrated a high-precision emission spectroscope using a chip-integrated microring resonator on thin-film lithium niobate, which measures the K lines from a HCL plasma source with a FWHM of 2 pm. We have also presented a temperature retrieval method to estimate the source temperature based on the precise linewidth measurement. Because of lithium niobate's wide transparency window, this device and temperature retrieval techniques could find broad applications in wildfire detection, atomic physics research, fusion reactor diagnostics, and astrophysics observations. With high resolution and chip integration, our device offers high performance, low cost, and robustness for airborne and spaceborne missions. 

Also, our technique allows the integration of multiple MRR fileters on the same chip, to simultaneously measure multiple emission lines. In this demonstration, we have measured the K doublet lines using a single chip but individually. In the future, one can couple the outputs of multiple MRRs each to a single photon detector and sweep their resonance all at the same time, for hyper-spectral measurement. Another prospect is to integrate other optical components on the same chip, such as WDM, quantum frequency converters, optical parametric amplifiers, and single photon detectors, to realize complex functions with exceptional SWaP parameters and operation robustness \cite{SNSPD1,SNSPD3,Hong_OPO,OPO_heater}.

\begin{acknowledgments}
YD thanks Zhaohui Ma and Zhan Li for training and helps on nanofabrication. This work was supported in part by NASA (Grant No. 80NSSC22K0286). Device fabrication was performed at Advanced Science Research Center (ASRC), City University of New York (CUNY), Columbia Nano Initiative (CNI), Columbia University, and Center for Nanoscale Systems (CNS), Harvard University.
\end{acknowledgments}

\appendix
\section{\label{appendix:fab}MRR Fabrication}
The MMR chip is fabricated from a commercial 300-nm thin film X-cut LiNbO$_3$ with 4.7-$\mu m$ SiO$_2$ and 0.5-mm Si. For lithography, we spin 800-nm HSQ resist on the sample. The MRR pattern is written by a 150 keV E-beam lithography tool (EBL, Elionix ELS-BODEN 150) with 1 nA current. The HSQ is then developed by TMAH for 30s and the sample is etched by an ion beam etching tool (IBE, INTLVAC Nanoquest) for 180 nm depth. We use RCA-I to clean the redeposition of LN and BOE 6:1 to clean the remaining HSQ. The sidewall angle ($82^{\circ}$) and top width are checked by a field emission scanning electron microscope (FE-SEM, Nova NanoSEM). Then we add 1.5-um SiO$_2$ cladding by plasma-enhanced chemical
vapor deposition (PECVD, PlasmaPro NGP80).

Next, PMMA 495A11 (in large thickness for easy lift-off) and PMMA 950A4 (for high resolution writing pattern) with a total thickness of 1.5 $\mu m$ is spun as a resist. A 50 keV EBL (Elionix ELS-LS50) is used to write the heater pattern. By using IPA:MIBK 1:1 as the resist developer, the pattern is created and 10/100 nm Ti/Pt is deposited by electron-beam deposition (E-beam evaporator, Angstrom Ultra High Vacuum Nexdep). Finally, the lift-off by Remover PG for 30 minutes removes the remaining resist and produces the zigzag metal heater pattern.
\section{\label{appendix:cali}Calibration Methods}
\begin{figure}[h!]
         \centering \includegraphics[width=0.49\textwidth]{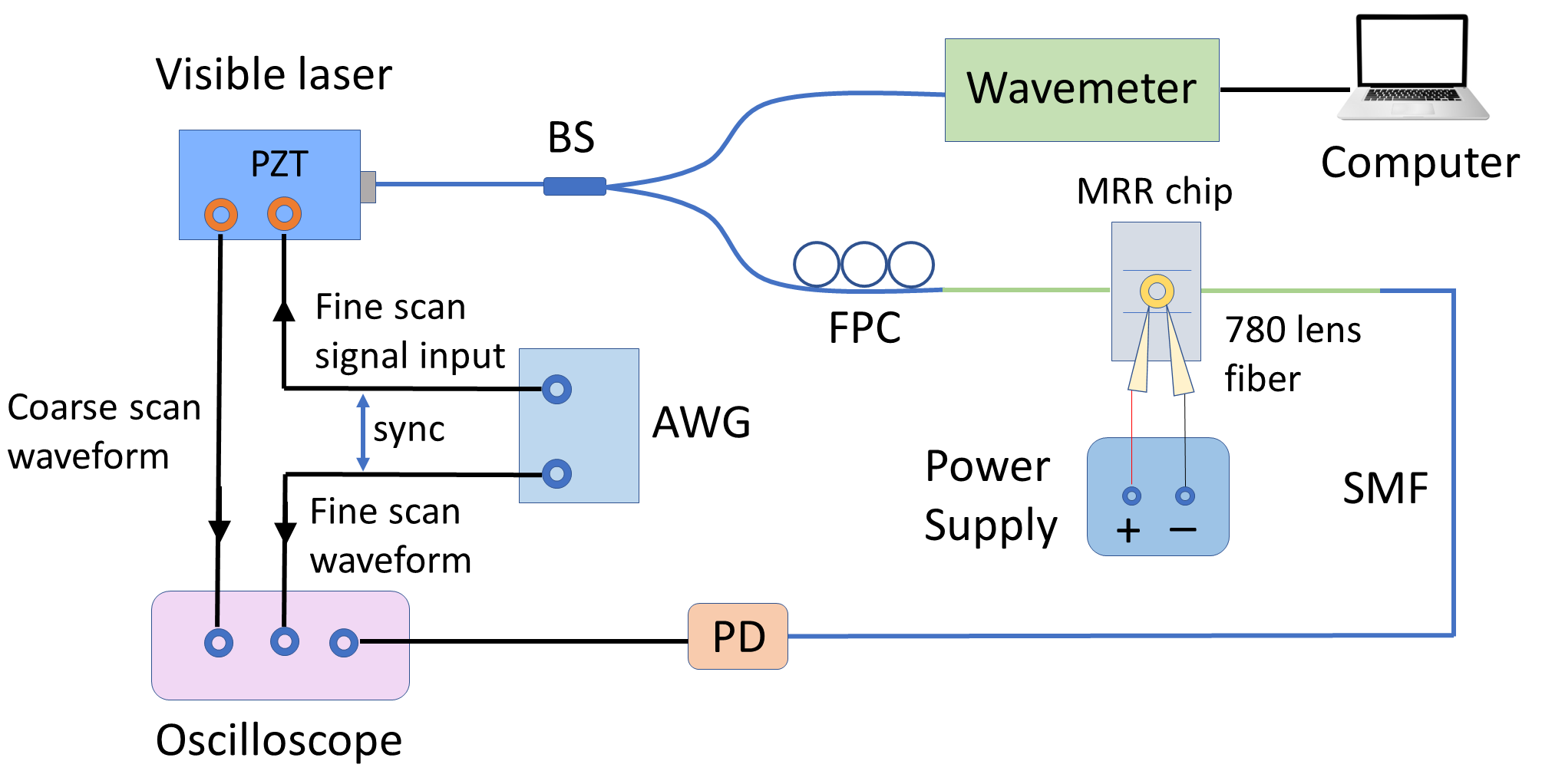}
    \caption{\justifying Calibration setup. PZT: piezoelectric transducer. BS: beam splitter. FPC: fiber polarization controller. PD: photodiode. AWG: arbitrary waveform generators. The blue, green, and black lines represent the SMF, the 780 nm lens fiber, and the electric wires, respectively. The coarse and fine scan waveforms are monitored by the oscilloscope.}
        \label{fig:setup2}
\end{figure}

The calibration system is shown in Fig.~\ref{fig:setup2}.
The calibration laser should have high resolution and minimal fluctuation. A tunable laser (New Focus TLB-6700) is used, with its output split by a beam splitter (BS). One output path is directed to a wavemeter (Msquared SolsTis) to monitor the wavelength scanning over time. The other path passes through a fiber polarization controller (FPC) and is coupled into the input port of the micro-ring resonator (MRR) using a 780 nm lens fiber (OZ OPTICS LTD). Light with a wavelength aligned to the resonator’s resonance wavelength is filtered and directed to the drop port. This filtered light is coupled into another 780 nm lens fiber, connected to a single-mode fiber (SMF), and detected by an amplified photodiode (PD, Thorlabs PDA100A2), which converts the optical signal to an electrical signal. The electrical signal is then monitored by an oscilloscope (Tektronix DPO 2004B) to find the resonance peaks. Simultaneously, the laser performs coarse wavelength scanning using an internal DC motor, with the wavelength monitored by the wavemeter and the corresponding electrical signals observed on the oscilloscope.

The fine scanning of the laser wavelength can be achieved through the external frequency modulation of a piezoelectric transducer (PZT) driver using channel 1 (CH1) of an arbitrary waveform generator (AWG, Rigol DG4162). The synchronized electrical signal from CH2 can be monitored via an oscilloscope.

\begin{figure}[h!]
         \centering \includegraphics[width=0.49\textwidth]{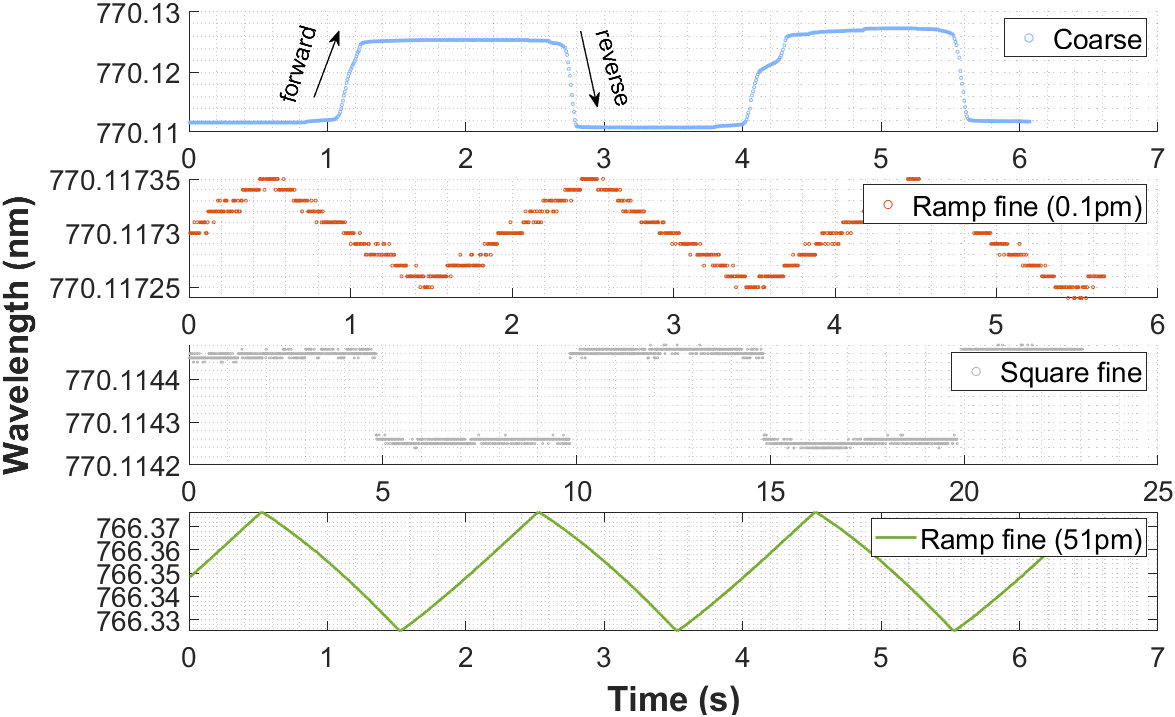}
    \caption{\justifying The wavemeter reading of the wavelength scan for laser coarse scan by an internal DC motor and fine scan by external frequency modulation of a PZT. The top subfigure shows the coarse scan, with 10-pm minimum scan range. For fine scan, ramp (0.1-pm minimum range) and square waveforms are applied to obtain laser step resolution and fluctuation, respectively. The bottom subfigure presents the fine scan with a ramp waveform applied (51-pm range) used to characterize the MRR linewidth.}
        \label{fig:Scan_All}
\end{figure}

To assess the resolution and stability of the laser, we compared its coarse and fine scanning capabilities as shown in Fig.~\ref{fig:Scan_All}. For coarse scan, we set a minimum scan range of 10 pm and a minimum scan speed of 10 pm/s. The minimum scan step resolution reaches 0.5 pm, but the fluctuation was as large as 1 pm in the forward scan and up to 2 pm in the reverse scan.

In the fine-scan mode, a ramp waveform with a peak-to-peak amplitude of 2 mV and a period of 2 s is applied from the AWG to the PZT, producing an output wavelength range of 0.1 pm. A square waveform from the AWG is also used as input signal to monitor the fluctuation of laser when the output wavelength remains constant. By analyzing the ramp and square scan data points, we determine that the laser resolution reaches 0.01 pm with a fluctuation of 0.03 pm.

Here, we also present the methods of characterizing the MRR chip. For coarse scan, a low laser input power -15 dBm is given to prevent skewing of the resonance peak due to optical bistability \cite{bistability,silicon}, which may give inaccurate linewidth measurements. The scan range is set to 2 nm with a speed of 1 pm/ms, starting at a wavelength of 766.1 nm. The time interval between starting time and the resonance peak time is measured on oscilloscope and converted to wavelength by the scan speed.

For the fine scanning, a peak-to-peak voltage of 2.2 V is applied to the laser PZT, resulting in a laser output wavelength range of 51 pm. The frequency of AWG is set to 500 mHz (2-second period), therefore the scanning speed is 0.051 pm/ms. By setting percent of the maximum PZT voltage to 50$\%$ we can adjust the fine scan range near the resonance wavelength located by the coarse scan, as shown in Fig.~\ref{fig:Scan_All}. Since the PD output voltage is linear to the incident light power, its FWHM can be found by directly reading the voltage at half of the maximum voltage value. By checking time interval on oscilloscope we can convert time to wavelength to obtain the resonance wavelength and linewidth precisely with accuracy $\pm$0.03 pm.

\section{\label{appendix:insight} Degree of Ionization and Broadening Mechanism Competition in HCL}
\begin{figure}[h!]
\captionsetup[subfloat]{position=top,justification=raggedright,singlelinecheck=false} 
\centering
\subfloat[][\label{fig:lamp_plasma}]{\includegraphics[width=\linewidth]{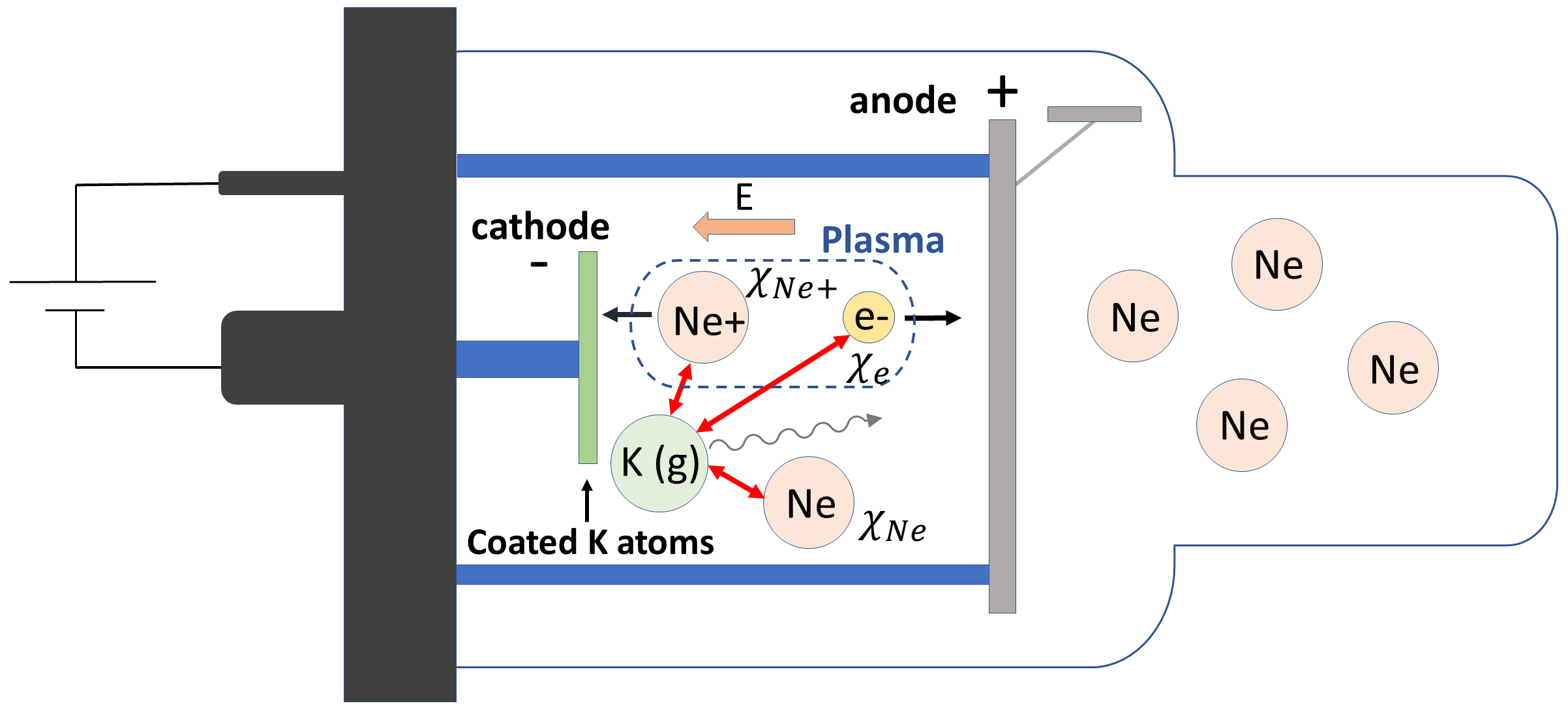}}\\
\subfloat[][\label{fig:ion1}]{\includegraphics[width=0.52\linewidth]{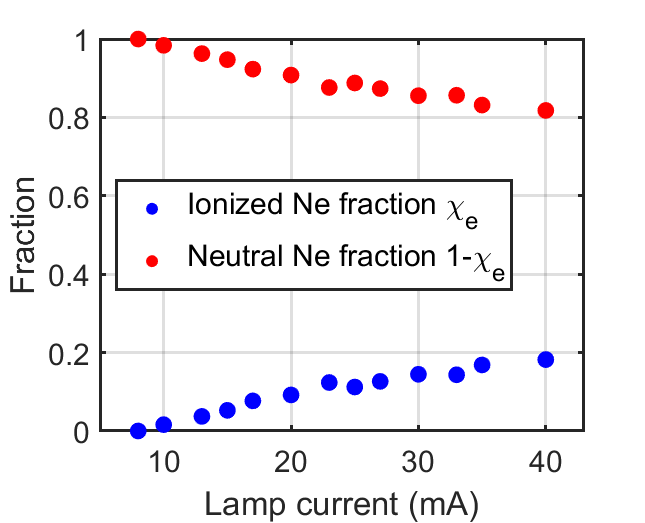}}
\subfloat[][\label{fig:ion2}]{\includegraphics[width=0.52\linewidth]{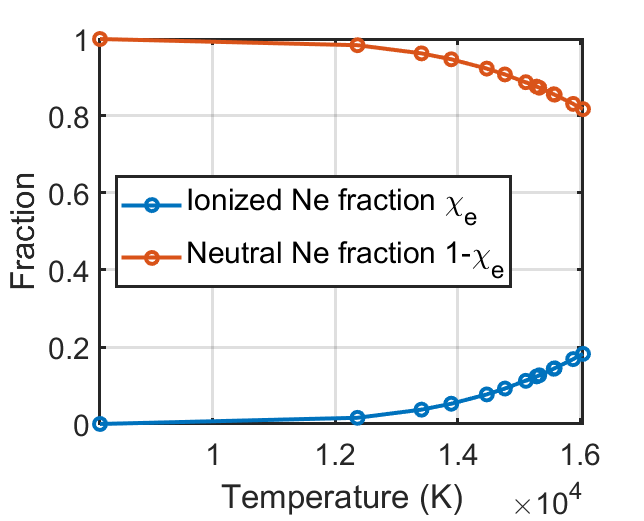}}\\
\subfloat[][\label{fig:ion3}]{\includegraphics[width=0.52\linewidth]{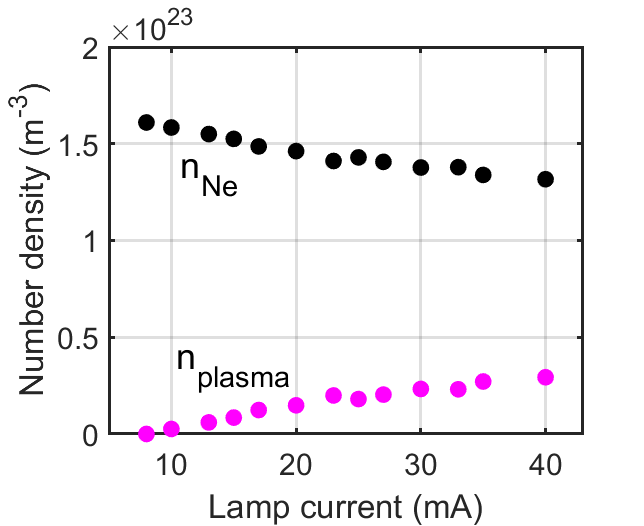}}
\subfloat[][\label{fig:ion4}]{\includegraphics[width=0.52\linewidth]{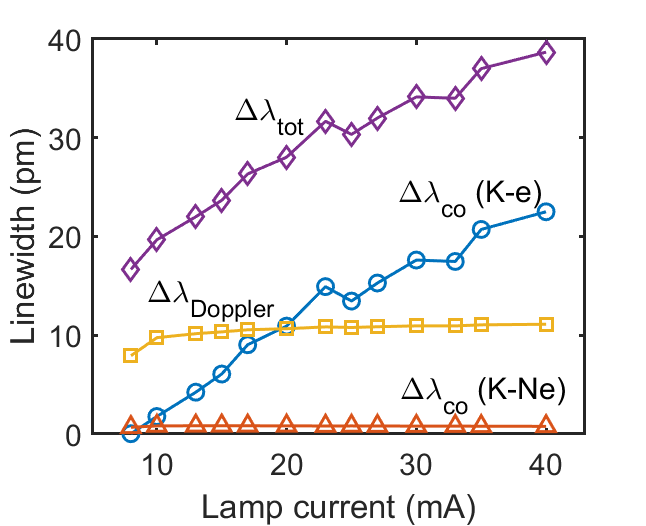}}
\caption{\justifying (a) An illustration of collision broadening mechanisms in the HCL. E denotes the electric field between HCL cathode and anode. K (g) denotes the potassium vapor. The wavy line represents the emitted K lines during the potassium decay. The red double-arrows represent the three collision processes between e-K, Ne$^+$-K, and Ne-K. Ionization fraction vs (b) lamp current and (c) retrieved temperature. (d) Plasma and neutral Neon density vs lamp current. (e) HCL broadening mechanisms comparison. $\Delta\lambda_{co}$: collision broadening linewidth. $\Delta\lambda_{tot}$: total broadening linewidth. }
\label{fig:ionization}
\end{figure}

Fig.~\subref*{fig:lamp_plasma} illustrates the formation of the conducting plasma and the collisions between potassium atoms and gas species in the HCL. An intensive electric field (denoted as E) is generated to ionize Ne inert gas between the cathode and anode by applying 280 V voltage and 8 mA current. The ionized Neon are attracted to the cathode side to bomber and vaporize the coated K atoms. The electrons emitted from the cathode, along with the ionized Neon, form a conducting plasma discharge. The excited potassium vapor, denoted as K (g), collides with the plasma and the neutral Neon atoms, leading to the emission of broadened K lines.

The degree of ionization $\chi_e$ and the fraction of unionized Ne $\chi_{Ne}$ versus lamp current, as well as ionization fraction vs retrieved temperature are plotted in Fig.~\subref*{fig:ion1} and Fig.~\subref*{fig:ion2}.
By using the fraction we obtained, the particle density vs lamp current are plotted in Fig.~\subref*{fig:ion3}, and the linewidth broadening mechanisms of the HCL can be compared in Fig.~\subref*{fig:ion4}. From the figure, it is evident that the collision broadening due to K-e collisions dominates over K-Ne collisions. Above specific current, it even dominates the doppler broadening. Collision broadening from K-Ne$^+$ is ignored here, because it has an even weaker contribution than K-Ne.

\section{\label{appendix:no plasma HCL}Retrieval Model for HCL without Plasma Incorporated}
\begin{figure}
         \centering \includegraphics[width=0.4\textwidth]{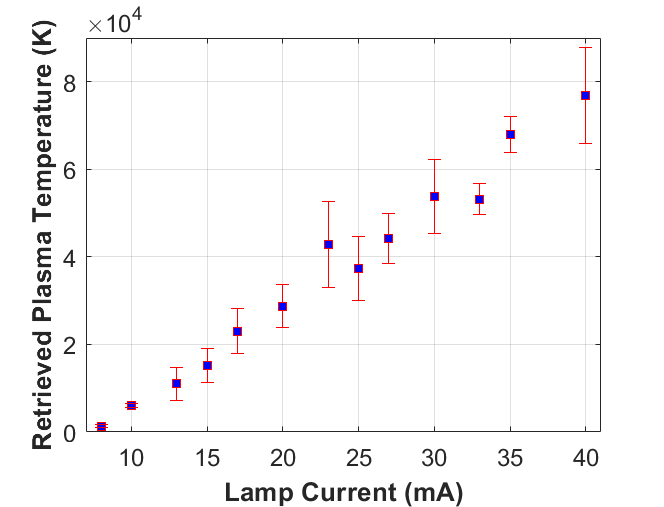}
    \caption{\justifying Retrieved temperature vs lamp current without plasma incorporated.}
        \label{fig:temp}
\end{figure}

Here we discuss a simpler model, where the plasmons formed by ionized Ne and electrons are ignored. Only unionized neutral Ne gas number density is considered here. Assuming ideal gas law condition is satisfied here, we use the same value of the gas particle density as the one used in main text $n=1.61\times 10^{23}$ m$^{-3}$, which remains constant due to the sealed HCL glass tube. Given each observed linewidth value, we are able to extract the corresponding temperatures by using the collision model in Ref.~\cite{Loudon} with Eqs.~\labelcref{eq:lo_sum}, ~\labelcref{eq:Dop}, and~\labelcref{eq:Olivero}. The retrieved temperature versus lamp current is shown in Fig.~\ref{fig:temp}.
\section{\label{appendix:app}Retrieval Models for Applications}
For wildfire, the potassium atoms in biomass burning undergo collisions with air molecules. The collision broadening linewidth equation is modified for mixture of species in air as
\begin{equation}   \Delta\lambda_{collision}=\Delta\lambda_0+\frac{2\sqrt{2}\lambda^2}{c}\frac{p}{kT}\sum_s \chi_s\sigma_{s-K} \Bar{v}_{s-K}
\end{equation}
where $\chi_s$ represent mole fraction of the species $s$ in air and s$-$K represents collision between species $s$ and potassium atom. Here the number density $n=p/kT$ of air is not fixed and varies with temperature.  On the other hand, constant atmosphere pressure $p\approx760$ torr is assumed here. The fraction of different species in air are $\chi_{N_2}=0.78$, $\chi_{Ar}=9.3\times 10^{-3}$, $\chi_{O_2}=0.21$, $\chi_{CO_2}=4.1\times 10^{-4}$, and $\chi_{Ne}=1.8\times 10^{-5}$. 

\begin{figure}
         \centering \includegraphics[width=0.42\textwidth]{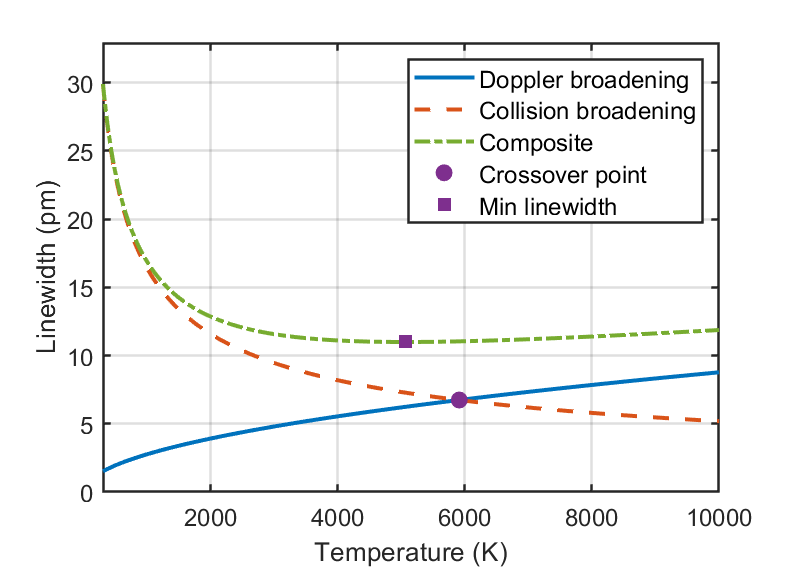}
    \caption{\justifying Competition between Doppler and collision broadening in wildfire. The crossover point of two broadening mechanisms is marked as the circular point and the minimum composited linewidth is marked as the square point.}
        \label{fig:wildfire}
\end{figure}

The Doppler and collision broadening linewidth versus temperature is shown in Fig.~\ref{fig:wildfire}.
\noindent The figure shows that below 5910 K (circular point), the collision broadening in wildfire play a more significant role than Doppler broadening. At 5070 K (square point), the composited broadening is at the minimum. Above or below this temperature will give broader linewidth. Flaming combustion in wildland fuels generally occurs at temperatures below 2500 K \cite{wildfire_temp2}, suggesting that collision broadening is the predominant mechanism in wildfire.

For fusion reactors, when neutral deuterium atoms are actively injected into the hot plasma, there is a probability that helium ions will capture an electron from these injected atoms, enabling them to emit line radiation in the visible wavelength region \cite{Tokamak_line_meas,plasma_diagnostics,Plasma_spectroscopy}. The collision broadening model is
\begin{equation}
    \Delta\lambda_{collision}=\Delta\lambda_0+\frac{2\sqrt{2}\lambda^2}{c}n \sum_s \chi_s\sigma_{s-M} \Bar{v}_{s-M}
\end{equation}
where M represents the radiating atom and $s$ is the particle that M collide with. $\chi_s$ is the fraction of the particle s, in which $\chi_e$ is the degree of ionization. If the total density is known, the model can be used to extract T. Conversely, if T is known, the plasma density can be retrieved. The plasma dynamics in Tokamak can be obtained by solving heat transport equation \cite{plasma_heat_eqn} in order to obtain $\chi_e$ for the collision broadening model.

In stellar bodies such as stars and AGN, it is essential to consider additional broadening mechanisms, such as rotational broadening and turbulence broadening \cite{Rotational_linewidth,AGN_turbo}. The observed profile becomes
\begin{equation}
    P_{observed}(\lambda)=G(\lambda)\otimes L(\lambda)\otimes X(\lambda)
\end{equation}
where $X(\lambda)$ refers to all the other broadening mechanisms besides Doppler and collision broadening.
The Stark broadening calculated by the semiclassical perturbation formalism (SCPF) \cite{AGN_collision} could be incorporated into the collision broadening of the retrieval model to account for the influence of electric fields from the charged particles in highly ionized stellar bodies.
\nocite{*}

\bibliography{citations}

\begin{thebibliography}{63}%
\makeatletter
\providecommand \@ifxundefined [1]{%
 \@ifx{#1\undefined}
}%
\providecommand \@ifnum [1]{%
 \ifnum #1\expandafter \@firstoftwo
 \else \expandafter \@secondoftwo
 \fi
}%
\providecommand \@ifx [1]{%
 \ifx #1\expandafter \@firstoftwo
 \else \expandafter \@secondoftwo
 \fi
}%
\providecommand \natexlab [1]{#1}%
\providecommand \enquote  [1]{``#1''}%
\providecommand \bibnamefont  [1]{#1}%
\providecommand \bibfnamefont [1]{#1}%
\providecommand \citenamefont [1]{#1}%
\providecommand \href@noop [0]{\@secondoftwo}%
\providecommand \href [0]{\begingroup \@sanitize@url \@href}%
\providecommand \@href[1]{\@@startlink{#1}\@@href}%
\providecommand \@@href[1]{\endgroup#1\@@endlink}%
\providecommand \@sanitize@url [0]{\catcode `\\12\catcode `\$12\catcode `\&12\catcode `\#12\catcode `\^12\catcode `\_12\catcode `\%12\relax}%
\providecommand \@@startlink[1]{}%
\providecommand \@@endlink[0]{}%
\providecommand \url  [0]{\begingroup\@sanitize@url \@url }%
\providecommand \@url [1]{\endgroup\@href {#1}{\urlprefix }}%
\providecommand \urlprefix  [0]{URL }%
\providecommand \Eprint [0]{\href }%
\providecommand \doibase [0]{https://doi.org/}%
\providecommand \selectlanguage [0]{\@gobble}%
\providecommand \bibinfo  [0]{\@secondoftwo}%
\providecommand \bibfield  [0]{\@secondoftwo}%
\providecommand \translation [1]{[#1]}%
\providecommand \BibitemOpen [0]{}%
\providecommand \bibitemStop [0]{}%
\providecommand \bibitemNoStop [0]{.\EOS\space}%
\providecommand \EOS [0]{\spacefactor3000\relax}%
\providecommand \BibitemShut  [1]{\csname bibitem#1\endcsname}%
\let\auto@bib@innerbib\@empty
\bibitem [{\citenamefont {Amici}\ \emph {et~al.}(2011)\citenamefont {Amici}, \citenamefont {Wooster},\ and\ \citenamefont {Piscini}}]{K_wildfire}%
  \BibitemOpen
  \bibfield  {author} {\bibinfo {author} {\bibfnamefont {S.}~\bibnamefont {Amici}}, \bibinfo {author} {\bibfnamefont {M.~J.}\ \bibnamefont {Wooster}},\ and\ \bibinfo {author} {\bibfnamefont {A.}~\bibnamefont {Piscini}},\ }\bibfield  {title} {\bibinfo {title} {Multi-resolution spectral analysis of wildfire potassium emission signatures using laboratory, airborne and spaceborne remote sensing},\ }\href {https://doi.org/https://doi.org/10.1016/j.rse.2011.02.022} {\bibfield  {journal} {\bibinfo  {journal} {Remote Sensing of Environment}\ }\textbf {\bibinfo {volume} {115}},\ \bibinfo {pages} {1811} (\bibinfo {year} {2011})}\BibitemShut {NoStop}%
\bibitem [{\citenamefont {Bonaventura}\ \emph {et~al.}(2023)\citenamefont {Bonaventura}, \citenamefont {Jakobsen}, \citenamefont {Ferruit}, \citenamefont {Arribas},\ and\ \citenamefont {Giardino}}]{weight}%
  \BibitemOpen
  \bibfield  {author} {\bibinfo {author} {\bibfnamefont {N.}~\bibnamefont {Bonaventura}}, \bibinfo {author} {\bibfnamefont {P.}~\bibnamefont {Jakobsen}}, \bibinfo {author} {\bibfnamefont {P.}~\bibnamefont {Ferruit}}, \bibinfo {author} {\bibfnamefont {S.}~\bibnamefont {Arribas}},\ and\ \bibinfo {author} {\bibfnamefont {G.}~\bibnamefont {Giardino}},\ }\bibfield  {title} {\bibinfo {title} {The {Near}-{Infrared} {Spectrograph} ({NIRSpec}) on the \textit{{James} {Webb}} {Space} {Telescope}: {V}. {Optimal} algorithms for planning multi-object spectroscopic observations},\ }\href {https://doi.org/10.1051/0004-6361/202245403} {\bibfield  {journal} {\bibinfo  {journal} {Astronomy \& Astrophysics}\ }\textbf {\bibinfo {volume} {672}},\ \bibinfo {pages} {A40} (\bibinfo {year} {2023})}\BibitemShut {NoStop}%
\bibitem [{\citenamefont {Greenhouse}(2016)}]{James_Webb}%
  \BibitemOpen
  \bibfield  {author} {\bibinfo {author} {\bibfnamefont {M.}~\bibnamefont {Greenhouse}},\ }\bibfield  {title} {\bibinfo {title} {The james webb space telescope: Mission overview and status},\ }in\ \href {https://doi.org/10.1109/AERO.2016.7500940} {\emph {\bibinfo {booktitle} {2016 IEEE Aerospace Conference}}}\ (\bibinfo {year} {2016})\ pp.\ \bibinfo {pages} {1--11}\BibitemShut {NoStop}%
\bibitem [{\citenamefont {Fantz}(2006)}]{plasma_diagnostics}%
  \BibitemOpen
  \bibfield  {author} {\bibinfo {author} {\bibfnamefont {U.}~\bibnamefont {Fantz}},\ }\bibfield  {title} {\bibinfo {title} {Basics of plasma spectroscopy},\ }\href {https://doi.org/10.1088/0963-0252/15/4/S01} {\bibfield  {journal} {\bibinfo  {journal} {Plasma Sources Science and Technology}\ }\textbf {\bibinfo {volume} {15}},\ \bibinfo {pages} {S137} (\bibinfo {year} {2006})}\BibitemShut {NoStop}%
\bibitem [{\citenamefont {Sharma}\ \emph {et~al.}(2023)\citenamefont {Sharma}, \citenamefont {Roppel}, \citenamefont {Murphy}, \citenamefont {Beegle}, \citenamefont {Bhartia}, \citenamefont {Steele}, \citenamefont {Hollis}, \citenamefont {Siljeström}, \citenamefont {McCubbin}, \citenamefont {Asher}, \citenamefont {Abbey}, \citenamefont {Allwood}, \citenamefont {Berger}, \citenamefont {Bleefeld}, \citenamefont {Burton}, \citenamefont {Bykov}, \citenamefont {Cardarelli}, \citenamefont {Conrad}, \citenamefont {Corpolongo}, \citenamefont {Czaja}, \citenamefont {DeFlores}, \citenamefont {Edgett}, \citenamefont {Farley}, \citenamefont {Fornaro}, \citenamefont {Fox}, \citenamefont {Fries}, \citenamefont {Harker}, \citenamefont {Hickman-Lewis}, \citenamefont {Huggett}, \citenamefont {Imbeah}, \citenamefont {Jakubek}, \citenamefont {Kah}, \citenamefont {Lee}, \citenamefont {Liu}, \citenamefont {Magee}, \citenamefont {Minitti}, \citenamefont {Moore}, \citenamefont {Pascuzzo}, \citenamefont {Rodriguez
  Sanchez-Vahamonde}, \citenamefont {Scheller}, \citenamefont {Shkolyar}, \citenamefont {Stack}, \citenamefont {Steadman}, \citenamefont {Tuite}, \citenamefont {Uckert}, \citenamefont {Werynski}, \citenamefont {Wiens}, \citenamefont {Williams}, \citenamefont {Winchell}, \citenamefont {Kennedy},\ and\ \citenamefont {Yanchilina}}]{Mars}%
  \BibitemOpen
  \bibfield  {author} {\bibinfo {author} {\bibfnamefont {S.}~\bibnamefont {Sharma}}, \bibinfo {author} {\bibfnamefont {R.~D.}\ \bibnamefont {Roppel}}, \bibinfo {author} {\bibfnamefont {A.~E.}\ \bibnamefont {Murphy}}, \bibinfo {author} {\bibfnamefont {L.~W.}\ \bibnamefont {Beegle}}, \bibinfo {author} {\bibfnamefont {R.}~\bibnamefont {Bhartia}}, \bibinfo {author} {\bibfnamefont {A.}~\bibnamefont {Steele}}, \bibinfo {author} {\bibfnamefont {J.~R.}\ \bibnamefont {Hollis}}, \bibinfo {author} {\bibfnamefont {S.}~\bibnamefont {Siljeström}}, \bibinfo {author} {\bibfnamefont {F.~M.}\ \bibnamefont {McCubbin}}, \bibinfo {author} {\bibfnamefont {S.~A.}\ \bibnamefont {Asher}}, \bibinfo {author} {\bibfnamefont {W.~J.}\ \bibnamefont {Abbey}}, \bibinfo {author} {\bibfnamefont {A.~C.}\ \bibnamefont {Allwood}}, \bibinfo {author} {\bibfnamefont {E.~L.}\ \bibnamefont {Berger}}, \bibinfo {author} {\bibfnamefont {B.~L.}\ \bibnamefont {Bleefeld}}, \bibinfo {author} {\bibfnamefont {A.~S.}\ \bibnamefont {Burton}}, \bibinfo {author}
  {\bibfnamefont {S.~V.}\ \bibnamefont {Bykov}}, \bibinfo {author} {\bibfnamefont {E.~L.}\ \bibnamefont {Cardarelli}}, \bibinfo {author} {\bibfnamefont {P.~G.}\ \bibnamefont {Conrad}}, \bibinfo {author} {\bibfnamefont {A.}~\bibnamefont {Corpolongo}}, \bibinfo {author} {\bibfnamefont {A.~D.}\ \bibnamefont {Czaja}}, \bibinfo {author} {\bibfnamefont {L.~P.}\ \bibnamefont {DeFlores}}, \bibinfo {author} {\bibfnamefont {K.}~\bibnamefont {Edgett}}, \bibinfo {author} {\bibfnamefont {K.~A.}\ \bibnamefont {Farley}}, \bibinfo {author} {\bibfnamefont {T.}~\bibnamefont {Fornaro}}, \bibinfo {author} {\bibfnamefont {A.~C.}\ \bibnamefont {Fox}}, \bibinfo {author} {\bibfnamefont {M.~D.}\ \bibnamefont {Fries}}, \bibinfo {author} {\bibfnamefont {D.}~\bibnamefont {Harker}}, \bibinfo {author} {\bibfnamefont {K.}~\bibnamefont {Hickman-Lewis}}, \bibinfo {author} {\bibfnamefont {J.}~\bibnamefont {Huggett}}, \bibinfo {author} {\bibfnamefont {S.}~\bibnamefont {Imbeah}}, \bibinfo {author} {\bibfnamefont {R.~S.}\ \bibnamefont
  {Jakubek}}, \bibinfo {author} {\bibfnamefont {L.~C.}\ \bibnamefont {Kah}}, \bibinfo {author} {\bibfnamefont {C.}~\bibnamefont {Lee}}, \bibinfo {author} {\bibfnamefont {Y.}~\bibnamefont {Liu}}, \bibinfo {author} {\bibfnamefont {A.}~\bibnamefont {Magee}}, \bibinfo {author} {\bibfnamefont {M.}~\bibnamefont {Minitti}}, \bibinfo {author} {\bibfnamefont {K.~R.}\ \bibnamefont {Moore}}, \bibinfo {author} {\bibfnamefont {A.}~\bibnamefont {Pascuzzo}}, \bibinfo {author} {\bibfnamefont {C.}~\bibnamefont {Rodriguez Sanchez-Vahamonde}}, \bibinfo {author} {\bibfnamefont {E.~L.}\ \bibnamefont {Scheller}}, \bibinfo {author} {\bibfnamefont {S.}~\bibnamefont {Shkolyar}}, \bibinfo {author} {\bibfnamefont {K.~M.}\ \bibnamefont {Stack}}, \bibinfo {author} {\bibfnamefont {K.}~\bibnamefont {Steadman}}, \bibinfo {author} {\bibfnamefont {M.}~\bibnamefont {Tuite}}, \bibinfo {author} {\bibfnamefont {K.}~\bibnamefont {Uckert}}, \bibinfo {author} {\bibfnamefont {A.}~\bibnamefont {Werynski}}, \bibinfo {author} {\bibfnamefont {R.~C.}\
  \bibnamefont {Wiens}}, \bibinfo {author} {\bibfnamefont {A.~J.}\ \bibnamefont {Williams}}, \bibinfo {author} {\bibfnamefont {K.}~\bibnamefont {Winchell}}, \bibinfo {author} {\bibfnamefont {M.~R.}\ \bibnamefont {Kennedy}},\ and\ \bibinfo {author} {\bibfnamefont {A.}~\bibnamefont {Yanchilina}},\ }\bibfield  {title} {{\selectlanguage {english}\bibinfo {title} {Diverse organic-mineral associations in {Jezero} crater, {Mars}}},\ }\href {https://doi.org/10.1038/s41586-023-06143-z} {\bibfield  {journal} {\bibinfo  {journal} {Nature}\ }\textbf {\bibinfo {volume} {619}},\ \bibinfo {pages} {724} (\bibinfo {year} {2023})}\BibitemShut {NoStop}%
\bibitem [{\citenamefont {Morton}(2003)}]{NIST_K}%
  \BibitemOpen
  \bibfield  {author} {\bibinfo {author} {\bibfnamefont {D.~C.}\ \bibnamefont {Morton}},\ }\bibfield  {title} {{\selectlanguage {english}\bibinfo {title} {Atomic {Data} for {Resonance} {Absorption} {Lines}. {III}. {Wavelengths} {Longward} of the {Lyman} {Limit} for the {Elements} {Hydrogen} to {Gallium}}},\ }\href {https://doi.org/10.1086/377639} {\bibfield  {journal} {\bibinfo  {journal} {The Astrophysical Journal Supplement Series}\ }\textbf {\bibinfo {volume} {149}},\ \bibinfo {pages} {205} (\bibinfo {year} {2003})}\BibitemShut {NoStop}%
\bibitem [{\citenamefont {Falke}\ \emph {et~al.}(2006)\citenamefont {Falke}, \citenamefont {Tiemann}, \citenamefont {Lisdat}, \citenamefont {Schnatz},\ and\ \citenamefont {Grosche}}]{D_doublet_accurate}%
  \BibitemOpen
  \bibfield  {author} {\bibinfo {author} {\bibfnamefont {S.}~\bibnamefont {Falke}}, \bibinfo {author} {\bibfnamefont {E.}~\bibnamefont {Tiemann}}, \bibinfo {author} {\bibfnamefont {C.}~\bibnamefont {Lisdat}}, \bibinfo {author} {\bibfnamefont {H.}~\bibnamefont {Schnatz}},\ and\ \bibinfo {author} {\bibfnamefont {G.}~\bibnamefont {Grosche}},\ }\bibfield  {title} {\bibinfo {title} {Transition frequencies of the $d$ lines of $^{39}\mathrm{K}$, $^{40}\mathrm{K}$, and $^{41}\mathrm{K}$ measured with a femtosecond laser frequency comb},\ }\href {https://doi.org/10.1103/PhysRevA.74.032503} {\bibfield  {journal} {\bibinfo  {journal} {Phys. Rev. A}\ }\textbf {\bibinfo {volume} {74}},\ \bibinfo {pages} {032503} (\bibinfo {year} {2006})}\BibitemShut {NoStop}%
\bibitem [{\citenamefont {Vodacek}\ \emph {et~al.}(2002)\citenamefont {Vodacek}, \citenamefont {Kremens}, \citenamefont {Fordham}, \citenamefont {Vangorden}, \citenamefont {Luisi}, \citenamefont {Schott},\ and\ \citenamefont {Latham}}]{K_trans}%
  \BibitemOpen
  \bibfield  {author} {\bibinfo {author} {\bibfnamefont {A.}~\bibnamefont {Vodacek}}, \bibinfo {author} {\bibfnamefont {R.~L.}\ \bibnamefont {Kremens}}, \bibinfo {author} {\bibfnamefont {A.~J.}\ \bibnamefont {Fordham}}, \bibinfo {author} {\bibfnamefont {S.~C.}\ \bibnamefont {Vangorden}}, \bibinfo {author} {\bibfnamefont {D.}~\bibnamefont {Luisi}}, \bibinfo {author} {\bibfnamefont {J.~R.}\ \bibnamefont {Schott}},\ and\ \bibinfo {author} {\bibfnamefont {D.~J.}\ \bibnamefont {Latham}},\ }\bibfield  {title} {{\selectlanguage {english}\bibinfo {title} {Remote optical detection of biomass burning using a potassium emission signature}},\ }\href {https://doi.org/10.1080/01431160110109633} {\bibfield  {journal} {\bibinfo  {journal} {International Journal of Remote Sensing}\ }\textbf {\bibinfo {volume} {23}},\ \bibinfo {pages} {2721} (\bibinfo {year} {2002})}\BibitemShut {NoStop}%
\bibitem [{\citenamefont {Waigl}\ \emph {et~al.}(2019)\citenamefont {Waigl}, \citenamefont {Prakash}, \citenamefont {Stuefer}, \citenamefont {Verbyla},\ and\ \citenamefont {Dennison}}]{Alaska_K}%
  \BibitemOpen
  \bibfield  {author} {\bibinfo {author} {\bibfnamefont {C.~F.}\ \bibnamefont {Waigl}}, \bibinfo {author} {\bibfnamefont {A.}~\bibnamefont {Prakash}}, \bibinfo {author} {\bibfnamefont {M.}~\bibnamefont {Stuefer}}, \bibinfo {author} {\bibfnamefont {D.}~\bibnamefont {Verbyla}},\ and\ \bibinfo {author} {\bibfnamefont {P.}~\bibnamefont {Dennison}},\ }\bibfield  {title} {{\selectlanguage {english}\bibinfo {title} {Fire detection and temperature retrieval using {EO}-1 {Hyperion} data over selected {Alaskan} boreal forest fires}},\ }\href {https://doi.org/10.1016/j.jag.2019.03.004} {\bibfield  {journal} {\bibinfo  {journal} {International Journal of Applied Earth Observation and Geoinformation}\ }\textbf {\bibinfo {volume} {81}},\ \bibinfo {pages} {72} (\bibinfo {year} {2019})}\BibitemShut {NoStop}%
\bibitem [{\citenamefont {Dennison}\ \emph {et~al.}(2006)\citenamefont {Dennison}, \citenamefont {Charoensiri}, \citenamefont {Roberts}, \citenamefont {Peterson},\ and\ \citenamefont {Green}}]{wildfire_temp2}%
  \BibitemOpen
  \bibfield  {author} {\bibinfo {author} {\bibfnamefont {P.}~\bibnamefont {Dennison}}, \bibinfo {author} {\bibfnamefont {K.}~\bibnamefont {Charoensiri}}, \bibinfo {author} {\bibfnamefont {D.}~\bibnamefont {Roberts}}, \bibinfo {author} {\bibfnamefont {S.}~\bibnamefont {Peterson}},\ and\ \bibinfo {author} {\bibfnamefont {R.}~\bibnamefont {Green}},\ }\bibfield  {title} {{\selectlanguage {english}\bibinfo {title} {Wildfire temperature and land cover modeling using hyperspectral data}},\ }\href {https://doi.org/10.1016/j.rse.2005.10.007} {\bibfield  {journal} {\bibinfo  {journal} {Remote Sensing of Environment}\ }\textbf {\bibinfo {volume} {100}},\ \bibinfo {pages} {212} (\bibinfo {year} {2006})}\BibitemShut {NoStop}%
\bibitem [{\citenamefont {Tipler}\ and\ \citenamefont {Llewellyn}(2012)}]{Modern_physics}%
  \BibitemOpen
  \bibfield  {author} {\bibinfo {author} {\bibfnamefont {P.~A.}\ \bibnamefont {Tipler}}\ and\ \bibinfo {author} {\bibfnamefont {R.~A.}\ \bibnamefont {Llewellyn}},\ }\href@noop {} {\emph {\bibinfo {title} {Modern physics}}},\ \bibinfo {edition} {6th}\ ed.\ (\bibinfo  {publisher} {W. H. Freeman and Co},\ \bibinfo {address} {New York},\ \bibinfo {year} {2012})\BibitemShut {NoStop}%
\bibitem [{\citenamefont {Ding}\ \emph {et~al.}(2024)\citenamefont {Ding}, \citenamefont {Garofalo}, \citenamefont {Wang}, \citenamefont {Weisberg}, \citenamefont {Li}, \citenamefont {Jian}, \citenamefont {Eldon}, \citenamefont {Victor}, \citenamefont {Marinoni}, \citenamefont {Hu}, \citenamefont {Carvalho}, \citenamefont {Odstrčil}, \citenamefont {Wang}, \citenamefont {Hyatt}, \citenamefont {Osborne}, \citenamefont {Gong}, \citenamefont {Qian}, \citenamefont {Huang}, \citenamefont {McClenaghan}, \citenamefont {Holcomb},\ and\ \citenamefont {Hanson}}]{Tokamak_confine}%
  \BibitemOpen
  \bibfield  {author} {\bibinfo {author} {\bibfnamefont {S.}~\bibnamefont {Ding}}, \bibinfo {author} {\bibfnamefont {A.~M.}\ \bibnamefont {Garofalo}}, \bibinfo {author} {\bibfnamefont {H.~Q.}\ \bibnamefont {Wang}}, \bibinfo {author} {\bibfnamefont {D.~B.}\ \bibnamefont {Weisberg}}, \bibinfo {author} {\bibfnamefont {Z.~Y.}\ \bibnamefont {Li}}, \bibinfo {author} {\bibfnamefont {X.}~\bibnamefont {Jian}}, \bibinfo {author} {\bibfnamefont {D.}~\bibnamefont {Eldon}}, \bibinfo {author} {\bibfnamefont {B.~S.}\ \bibnamefont {Victor}}, \bibinfo {author} {\bibfnamefont {A.}~\bibnamefont {Marinoni}}, \bibinfo {author} {\bibfnamefont {Q.~M.}\ \bibnamefont {Hu}}, \bibinfo {author} {\bibfnamefont {I.~S.}\ \bibnamefont {Carvalho}}, \bibinfo {author} {\bibfnamefont {T.}~\bibnamefont {Odstrčil}}, \bibinfo {author} {\bibfnamefont {L.}~\bibnamefont {Wang}}, \bibinfo {author} {\bibfnamefont {A.~W.}\ \bibnamefont {Hyatt}}, \bibinfo {author} {\bibfnamefont {T.~H.}\ \bibnamefont {Osborne}}, \bibinfo {author} {\bibfnamefont {X.~Z.}\
  \bibnamefont {Gong}}, \bibinfo {author} {\bibfnamefont {J.~P.}\ \bibnamefont {Qian}}, \bibinfo {author} {\bibfnamefont {J.}~\bibnamefont {Huang}}, \bibinfo {author} {\bibfnamefont {J.}~\bibnamefont {McClenaghan}}, \bibinfo {author} {\bibfnamefont {C.~T.}\ \bibnamefont {Holcomb}},\ and\ \bibinfo {author} {\bibfnamefont {J.~M.}\ \bibnamefont {Hanson}},\ }\bibfield  {title} {{\selectlanguage {english}\bibinfo {title} {A high-density and high-confinement tokamak plasma regime for fusion energy}},\ }\href {https://doi.org/10.1038/s41586-024-07313-3} {\bibfield  {journal} {\bibinfo  {journal} {Nature}\ }\textbf {\bibinfo {volume} {629}},\ \bibinfo {pages} {555} (\bibinfo {year} {2024})}\BibitemShut {NoStop}%
\bibitem [{\citenamefont {Jaspers}\ \emph {et~al.}(2012)\citenamefont {Jaspers}, \citenamefont {Scheffer}, \citenamefont {Kappatou}, \citenamefont {van~der Valk}, \citenamefont {Durkut}, \citenamefont {Snijders}, \citenamefont {Marchuk}, \citenamefont {Biel}, \citenamefont {Pokol}, \citenamefont {Erdei}, \citenamefont {Zoletnik},\ and\ \citenamefont {Dunai}}]{Tokamak_line_meas}%
  \BibitemOpen
  \bibfield  {author} {\bibinfo {author} {\bibfnamefont {R.~J.~E.}\ \bibnamefont {Jaspers}}, \bibinfo {author} {\bibfnamefont {M.}~\bibnamefont {Scheffer}}, \bibinfo {author} {\bibfnamefont {A.}~\bibnamefont {Kappatou}}, \bibinfo {author} {\bibfnamefont {N.~C.~J.}\ \bibnamefont {van~der Valk}}, \bibinfo {author} {\bibfnamefont {M.}~\bibnamefont {Durkut}}, \bibinfo {author} {\bibfnamefont {B.}~\bibnamefont {Snijders}}, \bibinfo {author} {\bibfnamefont {O.}~\bibnamefont {Marchuk}}, \bibinfo {author} {\bibfnamefont {W.}~\bibnamefont {Biel}}, \bibinfo {author} {\bibfnamefont {G.~I.}\ \bibnamefont {Pokol}}, \bibinfo {author} {\bibfnamefont {G.}~\bibnamefont {Erdei}}, \bibinfo {author} {\bibfnamefont {S.}~\bibnamefont {Zoletnik}},\ and\ \bibinfo {author} {\bibfnamefont {D.}~\bibnamefont {Dunai}},\ }\bibfield  {title} {\bibinfo {title} {{A high etendue spectrometer suitable for core charge eXchange recombination spectroscopy on ITERa)}},\ }\href {https://doi.org/10.1063/1.4732058} {\bibfield  {journal} {\bibinfo
  {journal} {Review of Scientific Instruments}\ }\textbf {\bibinfo {volume} {83}},\ \bibinfo {pages} {10D515} (\bibinfo {year} {2012})}\BibitemShut {NoStop}%
\bibitem [{\citenamefont {Zheng}\ \emph {et~al.}(2019)\citenamefont {Zheng}, \citenamefont {Zou}, \citenamefont {Cai}, \citenamefont {Song}, \citenamefont {Chin}, \citenamefont {Liu}, \citenamefont {Lin}, \citenamefont {Kwong},\ and\ \citenamefont {Liu}}]{spectrometer_heater}%
  \BibitemOpen
  \bibfield  {author} {\bibinfo {author} {\bibfnamefont {S.~N.}\ \bibnamefont {Zheng}}, \bibinfo {author} {\bibfnamefont {J.}~\bibnamefont {Zou}}, \bibinfo {author} {\bibfnamefont {H.}~\bibnamefont {Cai}}, \bibinfo {author} {\bibfnamefont {J.~F.}\ \bibnamefont {Song}}, \bibinfo {author} {\bibfnamefont {L.~K.}\ \bibnamefont {Chin}}, \bibinfo {author} {\bibfnamefont {P.~Y.}\ \bibnamefont {Liu}}, \bibinfo {author} {\bibfnamefont {Z.~P.}\ \bibnamefont {Lin}}, \bibinfo {author} {\bibfnamefont {D.~L.}\ \bibnamefont {Kwong}},\ and\ \bibinfo {author} {\bibfnamefont {A.~Q.}\ \bibnamefont {Liu}},\ }\bibfield  {title} {{\selectlanguage {english}\bibinfo {title} {Microring resonator-assisted {Fourier} transform spectrometer with enhanced resolution and large bandwidth in single chip solution}},\ }\href {https://doi.org/10.1038/s41467-019-10282-1} {\bibfield  {journal} {\bibinfo  {journal} {Nature Communications}\ }\textbf {\bibinfo {volume} {10}},\ \bibinfo {pages} {2349} (\bibinfo {year} {2019})}\BibitemShut {NoStop}%
\bibitem [{\citenamefont {Pohl}\ \emph {et~al.}(2020)\citenamefont {Pohl}, \citenamefont {Reig~Escalé}, \citenamefont {Madi}, \citenamefont {Kaufmann}, \citenamefont {Brotzer}, \citenamefont {Sergeyev}, \citenamefont {Guldimann}, \citenamefont {Giaccari}, \citenamefont {Alberti}, \citenamefont {Meier},\ and\ \citenamefont {Grange}}]{broadband_spectrometer_LN}%
  \BibitemOpen
  \bibfield  {author} {\bibinfo {author} {\bibfnamefont {D.}~\bibnamefont {Pohl}}, \bibinfo {author} {\bibfnamefont {M.}~\bibnamefont {Reig~Escalé}}, \bibinfo {author} {\bibfnamefont {M.}~\bibnamefont {Madi}}, \bibinfo {author} {\bibfnamefont {F.}~\bibnamefont {Kaufmann}}, \bibinfo {author} {\bibfnamefont {P.}~\bibnamefont {Brotzer}}, \bibinfo {author} {\bibfnamefont {A.}~\bibnamefont {Sergeyev}}, \bibinfo {author} {\bibfnamefont {B.}~\bibnamefont {Guldimann}}, \bibinfo {author} {\bibfnamefont {P.}~\bibnamefont {Giaccari}}, \bibinfo {author} {\bibfnamefont {E.}~\bibnamefont {Alberti}}, \bibinfo {author} {\bibfnamefont {U.}~\bibnamefont {Meier}},\ and\ \bibinfo {author} {\bibfnamefont {R.}~\bibnamefont {Grange}},\ }\bibfield  {title} {{\selectlanguage {english}\bibinfo {title} {An integrated broadband spectrometer on thin-film lithium niobate}},\ }\href {https://doi.org/10.1038/s41566-019-0529-9} {\bibfield  {journal} {\bibinfo  {journal} {Nature Photonics}\ }\textbf {\bibinfo {volume} {14}},\ \bibinfo
  {pages} {24} (\bibinfo {year} {2020})}\BibitemShut {NoStop}%
\bibitem [{\citenamefont {Yuan}\ \emph {et~al.}(2021)\citenamefont {Yuan}, \citenamefont {Naveh}, \citenamefont {Watanabe}, \citenamefont {Taniguchi},\ and\ \citenamefont {Xia}}]{phosphorus_spectrometer}%
  \BibitemOpen
  \bibfield  {author} {\bibinfo {author} {\bibfnamefont {S.}~\bibnamefont {Yuan}}, \bibinfo {author} {\bibfnamefont {D.}~\bibnamefont {Naveh}}, \bibinfo {author} {\bibfnamefont {K.}~\bibnamefont {Watanabe}}, \bibinfo {author} {\bibfnamefont {T.}~\bibnamefont {Taniguchi}},\ and\ \bibinfo {author} {\bibfnamefont {F.}~\bibnamefont {Xia}},\ }\bibfield  {title} {{\selectlanguage {english}\bibinfo {title} {A wavelength-scale black phosphorus spectrometer}},\ }\href {https://doi.org/10.1038/s41566-021-00787-x} {\bibfield  {journal} {\bibinfo  {journal} {Nature Photonics}\ }\textbf {\bibinfo {volume} {15}},\ \bibinfo {pages} {601} (\bibinfo {year} {2021})}\BibitemShut {NoStop}%
\bibitem [{\citenamefont {Grotevent}\ \emph {et~al.}(2023)\citenamefont {Grotevent}, \citenamefont {Yakunin}, \citenamefont {Bachmann}, \citenamefont {Romero}, \citenamefont {V{\'a}zquez~de Aldana}, \citenamefont {Madi}, \citenamefont {Calame}, \citenamefont {Kovalenko},\ and\ \citenamefont {Shorubalko}}]{compact_photodetector}%
  \BibitemOpen
  \bibfield  {author} {\bibinfo {author} {\bibfnamefont {M.~J.}\ \bibnamefont {Grotevent}}, \bibinfo {author} {\bibfnamefont {S.}~\bibnamefont {Yakunin}}, \bibinfo {author} {\bibfnamefont {D.}~\bibnamefont {Bachmann}}, \bibinfo {author} {\bibfnamefont {C.}~\bibnamefont {Romero}}, \bibinfo {author} {\bibfnamefont {J.~R.}\ \bibnamefont {V{\'a}zquez~de Aldana}}, \bibinfo {author} {\bibfnamefont {M.}~\bibnamefont {Madi}}, \bibinfo {author} {\bibfnamefont {M.}~\bibnamefont {Calame}}, \bibinfo {author} {\bibfnamefont {M.~V.}\ \bibnamefont {Kovalenko}},\ and\ \bibinfo {author} {\bibfnamefont {I.}~\bibnamefont {Shorubalko}},\ }\bibfield  {title} {\bibinfo {title} {Integrated photodetectors for compact fourier-transform waveguide spectrometers},\ }\href@noop {} {\bibfield  {journal} {\bibinfo  {journal} {Nature Photonics}\ }\textbf {\bibinfo {volume} {17}},\ \bibinfo {pages} {59} (\bibinfo {year} {2023})}\BibitemShut {NoStop}%
\bibitem [{\citenamefont {Zhang}\ \emph {et~al.}(2024)\citenamefont {Zhang}, \citenamefont {Yi}, \citenamefont {Liu}, \citenamefont {Hong}, \citenamefont {Wang}, \citenamefont {Cao}, \citenamefont {Shi},\ and\ \citenamefont {Dai}}]{silicon_spectrometer}%
  \BibitemOpen
  \bibfield  {author} {\bibinfo {author} {\bibfnamefont {L.}~\bibnamefont {Zhang}}, \bibinfo {author} {\bibfnamefont {X.}~\bibnamefont {Yi}}, \bibinfo {author} {\bibfnamefont {D.}~\bibnamefont {Liu}}, \bibinfo {author} {\bibfnamefont {S.}~\bibnamefont {Hong}}, \bibinfo {author} {\bibfnamefont {G.}~\bibnamefont {Wang}}, \bibinfo {author} {\bibfnamefont {H.}~\bibnamefont {Cao}}, \bibinfo {author} {\bibfnamefont {Y.}~\bibnamefont {Shi}},\ and\ \bibinfo {author} {\bibfnamefont {D.}~\bibnamefont {Dai}},\ }\bibfield  {title} {\bibinfo {title} {Silicon photonic spectrometer with multiple customized wavelength bands},\ }\href@noop {} {\bibfield  {journal} {\bibinfo  {journal} {Photonics Research}\ }\textbf {\bibinfo {volume} {12}},\ \bibinfo {pages} {1016} (\bibinfo {year} {2024})}\BibitemShut {NoStop}%
\bibitem [{\citenamefont {Graydon}(2018)}]{on_chip_spectro}%
  \BibitemOpen
  \bibfield  {author} {\bibinfo {author} {\bibfnamefont {O.}~\bibnamefont {Graydon}},\ }\bibfield  {title} {{\selectlanguage {english}\bibinfo {title} {On-chip spectroscopy}},\ }\href {https://doi.org/10.1038/s41566-018-0149-9} {\bibfield  {journal} {\bibinfo  {journal} {Nature Photonics}\ }\textbf {\bibinfo {volume} {12}},\ \bibinfo {pages} {189} (\bibinfo {year} {2018})}\BibitemShut {NoStop}%
\bibitem [{\citenamefont {Liapis}\ \emph {et~al.}(2016)\citenamefont {Liapis}, \citenamefont {Gao}, \citenamefont {Siddiqui}, \citenamefont {Shi},\ and\ \citenamefont {Boyd}}]{spectro_Boyd}%
  \BibitemOpen
  \bibfield  {author} {\bibinfo {author} {\bibfnamefont {A.~C.}\ \bibnamefont {Liapis}}, \bibinfo {author} {\bibfnamefont {B.}~\bibnamefont {Gao}}, \bibinfo {author} {\bibfnamefont {M.~R.}\ \bibnamefont {Siddiqui}}, \bibinfo {author} {\bibfnamefont {Z.}~\bibnamefont {Shi}},\ and\ \bibinfo {author} {\bibfnamefont {R.~W.}\ \bibnamefont {Boyd}},\ }\bibfield  {title} {{\selectlanguage {english}\bibinfo {title} {On-chip spectroscopy with thermally tuned high-{Q} photonic crystal cavities}},\ }\href {https://doi.org/10.1063/1.4939659} {\bibfield  {journal} {\bibinfo  {journal} {Applied Physics Letters}\ }\textbf {\bibinfo {volume} {108}},\ \bibinfo {pages} {021105} (\bibinfo {year} {2016})}\BibitemShut {NoStop}%
\bibitem [{\citenamefont {Zhu}\ \emph {et~al.}(2021)\citenamefont {Zhu}, \citenamefont {Shao}, \citenamefont {Yu}, \citenamefont {Cheng}, \citenamefont {Desiatov}, \citenamefont {Xin}, \citenamefont {Hu}, \citenamefont {Holzgrafe}, \citenamefont {Ghosh}, \citenamefont {Shams-Ansari}, \citenamefont {Puma}, \citenamefont {Sinclair}, \citenamefont {Reimer}, \citenamefont {Zhang},\ and\ \citenamefont {Lončar}}]{Metal_absorb2}%
  \BibitemOpen
  \bibfield  {author} {\bibinfo {author} {\bibfnamefont {D.}~\bibnamefont {Zhu}}, \bibinfo {author} {\bibfnamefont {L.}~\bibnamefont {Shao}}, \bibinfo {author} {\bibfnamefont {M.}~\bibnamefont {Yu}}, \bibinfo {author} {\bibfnamefont {R.}~\bibnamefont {Cheng}}, \bibinfo {author} {\bibfnamefont {B.}~\bibnamefont {Desiatov}}, \bibinfo {author} {\bibfnamefont {C.~J.}\ \bibnamefont {Xin}}, \bibinfo {author} {\bibfnamefont {Y.}~\bibnamefont {Hu}}, \bibinfo {author} {\bibfnamefont {J.}~\bibnamefont {Holzgrafe}}, \bibinfo {author} {\bibfnamefont {S.}~\bibnamefont {Ghosh}}, \bibinfo {author} {\bibfnamefont {A.}~\bibnamefont {Shams-Ansari}}, \bibinfo {author} {\bibfnamefont {E.}~\bibnamefont {Puma}}, \bibinfo {author} {\bibfnamefont {N.}~\bibnamefont {Sinclair}}, \bibinfo {author} {\bibfnamefont {C.}~\bibnamefont {Reimer}}, \bibinfo {author} {\bibfnamefont {M.}~\bibnamefont {Zhang}},\ and\ \bibinfo {author} {\bibfnamefont {M.}~\bibnamefont {Lončar}},\ }\bibfield  {title} {{\selectlanguage {english}\bibinfo {title}
  {Integrated photonics on thin-film lithium niobate}},\ }\href {https://doi.org/10.1364/AOP.411024} {\bibfield  {journal} {\bibinfo  {journal} {Advances in Optics and Photonics}\ }\textbf {\bibinfo {volume} {13}},\ \bibinfo {pages} {242} (\bibinfo {year} {2021})}\BibitemShut {NoStop}%
\bibitem [{\citenamefont {Boes}\ \emph {et~al.}(2023)\citenamefont {Boes}, \citenamefont {Chang}, \citenamefont {Langrock}, \citenamefont {Yu}, \citenamefont {Zhang}, \citenamefont {Lin}, \citenamefont {Lončar}, \citenamefont {Fejer}, \citenamefont {Bowers},\ and\ \citenamefont {Mitchell}}]{LN_science}%
  \BibitemOpen
  \bibfield  {author} {\bibinfo {author} {\bibfnamefont {A.}~\bibnamefont {Boes}}, \bibinfo {author} {\bibfnamefont {L.}~\bibnamefont {Chang}}, \bibinfo {author} {\bibfnamefont {C.}~\bibnamefont {Langrock}}, \bibinfo {author} {\bibfnamefont {M.}~\bibnamefont {Yu}}, \bibinfo {author} {\bibfnamefont {M.}~\bibnamefont {Zhang}}, \bibinfo {author} {\bibfnamefont {Q.}~\bibnamefont {Lin}}, \bibinfo {author} {\bibfnamefont {M.}~\bibnamefont {Lončar}}, \bibinfo {author} {\bibfnamefont {M.}~\bibnamefont {Fejer}}, \bibinfo {author} {\bibfnamefont {J.}~\bibnamefont {Bowers}},\ and\ \bibinfo {author} {\bibfnamefont {A.}~\bibnamefont {Mitchell}},\ }\bibfield  {title} {{\selectlanguage {english}\bibinfo {title} {Lithium niobate photonics: {Unlocking} the electromagnetic spectrum}},\ }\href {https://doi.org/10.1126/science.abj4396} {\bibfield  {journal} {\bibinfo  {journal} {Science}\ }\textbf {\bibinfo {volume} {379}},\ \bibinfo {pages} {eabj4396} (\bibinfo {year} {2023})}\BibitemShut {NoStop}%
\bibitem [{\citenamefont {Jin}\ \emph {et~al.}(2019)\citenamefont {Jin}, \citenamefont {Chen}, \citenamefont {Sua},\ and\ \citenamefont {Huang}}]{EOM}%
  \BibitemOpen
  \bibfield  {author} {\bibinfo {author} {\bibfnamefont {M.}~\bibnamefont {Jin}}, \bibinfo {author} {\bibfnamefont {J.-Y.}\ \bibnamefont {Chen}}, \bibinfo {author} {\bibfnamefont {Y.~M.}\ \bibnamefont {Sua}},\ and\ \bibinfo {author} {\bibfnamefont {Y.-P.}\ \bibnamefont {Huang}},\ }\bibfield  {title} {{\selectlanguage {english}\bibinfo {title} {High-extinction electro-optic modulation on lithium niobate thin film}},\ }\href {https://doi.org/10.1364/OL.44.001265} {\bibfield  {journal} {\bibinfo  {journal} {Optics Letters}\ }\textbf {\bibinfo {volume} {44}},\ \bibinfo {pages} {1265} (\bibinfo {year} {2019})}\BibitemShut {NoStop}%
\bibitem [{\citenamefont {Chen}\ \emph {et~al.}(2019)\citenamefont {Chen}, \citenamefont {Meng~Sua}, \citenamefont {Ma}, \citenamefont {Tang}, \citenamefont {Li},\ and\ \citenamefont {Huang}}]{chen}%
  \BibitemOpen
  \bibfield  {author} {\bibinfo {author} {\bibfnamefont {J.-y.}\ \bibnamefont {Chen}}, \bibinfo {author} {\bibfnamefont {Y.}~\bibnamefont {Meng~Sua}}, \bibinfo {author} {\bibfnamefont {Z.-h.}\ \bibnamefont {Ma}}, \bibinfo {author} {\bibfnamefont {C.}~\bibnamefont {Tang}}, \bibinfo {author} {\bibfnamefont {Z.}~\bibnamefont {Li}},\ and\ \bibinfo {author} {\bibfnamefont {Y.-p.}\ \bibnamefont {Huang}},\ }\bibfield  {title} {{\selectlanguage {english}\bibinfo {title} {Efficient parametric frequency conversion in lithium niobate nanophotonic chips}},\ }\href {https://doi.org/10.1364/OSAC.2.002914} {\bibfield  {journal} {\bibinfo  {journal} {OSA Continuum}\ }\textbf {\bibinfo {volume} {2}},\ \bibinfo {pages} {2914} (\bibinfo {year} {2019})}\BibitemShut {NoStop}%
\bibitem [{\citenamefont {Zhu}\ \emph {et~al.}(2024)\citenamefont {Zhu}, \citenamefont {Hu}, \citenamefont {Lu}, \citenamefont {Warner}, \citenamefont {Li}, \citenamefont {Song}, \citenamefont {Magalhães}, \citenamefont {Shams-Ansari}, \citenamefont {Cordaro}, \citenamefont {Sinclair},\ and\ \citenamefont {Lončar}}]{High_Q}%
  \BibitemOpen
  \bibfield  {author} {\bibinfo {author} {\bibfnamefont {X.}~\bibnamefont {Zhu}}, \bibinfo {author} {\bibfnamefont {Y.}~\bibnamefont {Hu}}, \bibinfo {author} {\bibfnamefont {S.}~\bibnamefont {Lu}}, \bibinfo {author} {\bibfnamefont {H.~K.}\ \bibnamefont {Warner}}, \bibinfo {author} {\bibfnamefont {X.}~\bibnamefont {Li}}, \bibinfo {author} {\bibfnamefont {Y.}~\bibnamefont {Song}}, \bibinfo {author} {\bibfnamefont {L.}~\bibnamefont {Magalhães}}, \bibinfo {author} {\bibfnamefont {A.}~\bibnamefont {Shams-Ansari}}, \bibinfo {author} {\bibfnamefont {A.}~\bibnamefont {Cordaro}}, \bibinfo {author} {\bibfnamefont {N.}~\bibnamefont {Sinclair}},\ and\ \bibinfo {author} {\bibfnamefont {M.}~\bibnamefont {Lončar}},\ }\href {https://doi.org/10.1364/PRJ.521172} {\bibfield  {journal} {\bibinfo  {journal} {Photonics Research}\ }\textbf {\bibinfo {volume} {12}},\ \bibinfo {pages} {A63} (\bibinfo {year} {2024})}\BibitemShut {NoStop}%
\bibitem [{\citenamefont {Zhang}\ \emph {et~al.}(2021)\citenamefont {Zhang}, \citenamefont {Sua}, \citenamefont {Chen}, \citenamefont {Ramanathan}, \citenamefont {Tang}, \citenamefont {Li}, \citenamefont {Hu},\ and\ \citenamefont {Huang}}]{CO2}%
  \BibitemOpen
  \bibfield  {author} {\bibinfo {author} {\bibfnamefont {J.}~\bibnamefont {Zhang}}, \bibinfo {author} {\bibfnamefont {Y.~M.}\ \bibnamefont {Sua}}, \bibinfo {author} {\bibfnamefont {J.-Y.}\ \bibnamefont {Chen}}, \bibinfo {author} {\bibfnamefont {J.}~\bibnamefont {Ramanathan}}, \bibinfo {author} {\bibfnamefont {C.}~\bibnamefont {Tang}}, \bibinfo {author} {\bibfnamefont {Z.}~\bibnamefont {Li}}, \bibinfo {author} {\bibfnamefont {Y.}~\bibnamefont {Hu}},\ and\ \bibinfo {author} {\bibfnamefont {Y.-P.}\ \bibnamefont {Huang}},\ }\bibfield  {title} {\bibinfo {title} {Carbon-dioxide absorption spectroscopy with solar photon counting and integrated lithium niobate micro-ring resonator},\ }\href {https://doi.org/10.1063/5.0045869} {\bibfield  {journal} {\bibinfo  {journal} {Applied Physics Letters}\ }\textbf {\bibinfo {volume} {118}},\ \bibinfo {pages} {171103} (\bibinfo {year} {2021})}\BibitemShut {NoStop}%
\bibitem [{\citenamefont {Zhang}\ \emph {et~al.}(2022)\citenamefont {Zhang}, \citenamefont {Sua}, \citenamefont {Hu}, \citenamefont {Ramanathan},\ and\ \citenamefont {Huang}}]{O2}%
  \BibitemOpen
  \bibfield  {author} {\bibinfo {author} {\bibfnamefont {J.}~\bibnamefont {Zhang}}, \bibinfo {author} {\bibfnamefont {Y.~M.}\ \bibnamefont {Sua}}, \bibinfo {author} {\bibfnamefont {Y.}~\bibnamefont {Hu}}, \bibinfo {author} {\bibfnamefont {J.}~\bibnamefont {Ramanathan}},\ and\ \bibinfo {author} {\bibfnamefont {Y.-P.}\ \bibnamefont {Huang}},\ }\bibfield  {title} {\bibinfo {title} {Oxygen a-band absorption spectroscopy with solar photon counting and lithium niobate nanophotonic circuits},\ }\bibfield  {journal} {\bibinfo  {journal} {Frontiers in Remote Sensing}\ }\textbf {\bibinfo {volume} {3}},\ \href {https://doi.org/10.3389/frsen.2022.1064244} {10.3389/frsen.2022.1064244} (\bibinfo {year} {2022})\BibitemShut {NoStop}%
\bibitem [{\citenamefont {Sarmiento}\ \emph {et~al.}(2018)\citenamefont {Sarmiento}, \citenamefont {Reiners}, \citenamefont {Huke}, \citenamefont {Bauer}, \citenamefont {Guenter}, \citenamefont {Seemann},\ and\ \citenamefont {Wolter}}]{white_dwarf}%
  \BibitemOpen
  \bibfield  {author} {\bibinfo {author} {\bibfnamefont {L.~F.}\ \bibnamefont {Sarmiento}}, \bibinfo {author} {\bibfnamefont {A.}~\bibnamefont {Reiners}}, \bibinfo {author} {\bibfnamefont {P.}~\bibnamefont {Huke}}, \bibinfo {author} {\bibfnamefont {F.~F.}\ \bibnamefont {Bauer}}, \bibinfo {author} {\bibfnamefont {E.~W.}\ \bibnamefont {Guenter}}, \bibinfo {author} {\bibfnamefont {U.}~\bibnamefont {Seemann}},\ and\ \bibinfo {author} {\bibfnamefont {U.}~\bibnamefont {Wolter}},\ }\bibfield  {title} {\bibinfo {title} {Comparing the emission spectra of {U} and {Th} hollow cathode lamps and a new {U} line list},\ }\href {https://doi.org/10.1051/0004-6361/201832871} {\bibfield  {journal} {\bibinfo  {journal} {Astronomy \& Astrophysics}\ }\textbf {\bibinfo {volume} {618}},\ \bibinfo {pages} {A118} (\bibinfo {year} {2018})}\BibitemShut {NoStop}%
\bibitem [{\citenamefont {Srećković}\ \emph {et~al.}(2001)\citenamefont {Srećković}, \citenamefont {Dimitrijević},\ and\ \citenamefont {Djeniže}}]{AGN_collision}%
  \BibitemOpen
  \bibfield  {author} {\bibinfo {author} {\bibfnamefont {A.}~\bibnamefont {Srećković}}, \bibinfo {author} {\bibfnamefont {M.~S.}\ \bibnamefont {Dimitrijević}},\ and\ \bibinfo {author} {\bibfnamefont {S.}~\bibnamefont {Djeniže}},\ }\bibfield  {title} {\bibinfo {title} {Stark broadening in {O} {III} spectrum},\ }\href {https://doi.org/10.1051/0004-6361:20010343} {\bibfield  {journal} {\bibinfo  {journal} {Astronomy \& Astrophysics}\ }\textbf {\bibinfo {volume} {371}},\ \bibinfo {pages} {354} (\bibinfo {year} {2001})}\BibitemShut {NoStop}%
\bibitem [{\citenamefont {Bogaerts}\ \emph {et~al.}(2012)\citenamefont {Bogaerts}, \citenamefont {De~Heyn}, \citenamefont {Van~Vaerenbergh}, \citenamefont {De~Vos}, \citenamefont {Kumar~Selvaraja}, \citenamefont {Claes}, \citenamefont {Dumon}, \citenamefont {Bienstman}, \citenamefont {Van~Thourhout},\ and\ \citenamefont {Baets}}]{silicon}%
  \BibitemOpen
  \bibfield  {author} {\bibinfo {author} {\bibfnamefont {W.}~\bibnamefont {Bogaerts}}, \bibinfo {author} {\bibfnamefont {P.}~\bibnamefont {De~Heyn}}, \bibinfo {author} {\bibfnamefont {T.}~\bibnamefont {Van~Vaerenbergh}}, \bibinfo {author} {\bibfnamefont {K.}~\bibnamefont {De~Vos}}, \bibinfo {author} {\bibfnamefont {S.}~\bibnamefont {Kumar~Selvaraja}}, \bibinfo {author} {\bibfnamefont {T.}~\bibnamefont {Claes}}, \bibinfo {author} {\bibfnamefont {P.}~\bibnamefont {Dumon}}, \bibinfo {author} {\bibfnamefont {P.}~\bibnamefont {Bienstman}}, \bibinfo {author} {\bibfnamefont {D.}~\bibnamefont {Van~Thourhout}},\ and\ \bibinfo {author} {\bibfnamefont {R.}~\bibnamefont {Baets}},\ }\bibfield  {title} {{\selectlanguage {english}\bibinfo {title} {Silicon microring resonators}},\ }\href {https://doi.org/10.1002/lpor.201100017} {\bibfield  {journal} {\bibinfo  {journal} {Laser \& Photonics Reviews}\ }\textbf {\bibinfo {volume} {6}},\ \bibinfo {pages} {47} (\bibinfo {year} {2012})}\BibitemShut {NoStop}%
\bibitem [{\citenamefont {Pozar}(2012)}]{Metal_absorb1}%
  \BibitemOpen
  \bibfield  {author} {\bibinfo {author} {\bibfnamefont {D.~M.}\ \bibnamefont {Pozar}},\ }\href@noop {} {{\selectlanguage {english}\emph {\bibinfo {title} {Microwave engineering}}}},\ \bibinfo {edition} {fourth edition}\ ed.\ (\bibinfo  {publisher} {John Wiley \& Sons, Inc},\ \bibinfo {address} {Hoboken, NJ},\ \bibinfo {year} {2012})\BibitemShut {NoStop}%
\bibitem [{\citenamefont {Boyajian}\ \emph {et~al.}(2012)\citenamefont {Boyajian}, \citenamefont {McAlister}, \citenamefont {Van~Belle}, \citenamefont {Gies}, \citenamefont {Ten~Brummelaar}, \citenamefont {Von~Braun}, \citenamefont {Farrington}, \citenamefont {Goldfinger}, \citenamefont {O'Brien}, \citenamefont {Parks}, \citenamefont {Richardson}, \citenamefont {Ridgway}, \citenamefont {Schaefer}, \citenamefont {Sturmann}, \citenamefont {Sturmann}, \citenamefont {Touhami}, \citenamefont {Turner},\ and\ \citenamefont {White}}]{star_temp1}%
  \BibitemOpen
  \bibfield  {author} {\bibinfo {author} {\bibfnamefont {T.~S.}\ \bibnamefont {Boyajian}}, \bibinfo {author} {\bibfnamefont {H.~A.}\ \bibnamefont {McAlister}}, \bibinfo {author} {\bibfnamefont {G.}~\bibnamefont {Van~Belle}}, \bibinfo {author} {\bibfnamefont {D.~R.}\ \bibnamefont {Gies}}, \bibinfo {author} {\bibfnamefont {T.~A.}\ \bibnamefont {Ten~Brummelaar}}, \bibinfo {author} {\bibfnamefont {K.}~\bibnamefont {Von~Braun}}, \bibinfo {author} {\bibfnamefont {C.}~\bibnamefont {Farrington}}, \bibinfo {author} {\bibfnamefont {P.~J.}\ \bibnamefont {Goldfinger}}, \bibinfo {author} {\bibfnamefont {D.}~\bibnamefont {O'Brien}}, \bibinfo {author} {\bibfnamefont {J.~R.}\ \bibnamefont {Parks}}, \bibinfo {author} {\bibfnamefont {N.~D.}\ \bibnamefont {Richardson}}, \bibinfo {author} {\bibfnamefont {S.}~\bibnamefont {Ridgway}}, \bibinfo {author} {\bibfnamefont {G.}~\bibnamefont {Schaefer}}, \bibinfo {author} {\bibfnamefont {L.}~\bibnamefont {Sturmann}}, \bibinfo {author} {\bibfnamefont {J.}~\bibnamefont {Sturmann}},
  \bibinfo {author} {\bibfnamefont {Y.}~\bibnamefont {Touhami}}, \bibinfo {author} {\bibfnamefont {N.~H.}\ \bibnamefont {Turner}},\ and\ \bibinfo {author} {\bibfnamefont {R.}~\bibnamefont {White}},\ }\bibfield  {title} {\bibinfo {title} {{STELLAR} {DIAMETERS} {AND} {TEMPERATURES}. {I}. {MAIN}-{SEQUENCE} {A}, {F}, {AND} {G} {STARS}},\ }\href {https://doi.org/10.1088/0004-637X/746/1/101} {\bibfield  {journal} {\bibinfo  {journal} {The Astrophysical Journal}\ }\textbf {\bibinfo {volume} {746}},\ \bibinfo {pages} {101} (\bibinfo {year} {2012})}\BibitemShut {NoStop}%
\bibitem [{\citenamefont {Goebel}\ \emph {et~al.}(2021)\citenamefont {Goebel}, \citenamefont {Becatti}, \citenamefont {Mikellides},\ and\ \citenamefont {Lopez~Ortega}}]{plasma_temp_HCL}%
  \BibitemOpen
  \bibfield  {author} {\bibinfo {author} {\bibfnamefont {D.~M.}\ \bibnamefont {Goebel}}, \bibinfo {author} {\bibfnamefont {G.}~\bibnamefont {Becatti}}, \bibinfo {author} {\bibfnamefont {I.~G.}\ \bibnamefont {Mikellides}},\ and\ \bibinfo {author} {\bibfnamefont {A.}~\bibnamefont {Lopez~Ortega}},\ }\bibfield  {title} {{\selectlanguage {english}\bibinfo {title} {Plasma hollow cathodes}},\ }\href {https://doi.org/10.1063/5.0051228} {\bibfield  {journal} {\bibinfo  {journal} {Journal of Applied Physics}\ }\textbf {\bibinfo {volume} {130}},\ \bibinfo {pages} {050902} (\bibinfo {year} {2021})}\BibitemShut {NoStop}%
\bibitem [{\citenamefont {Loudon}(2000)}]{Loudon}%
  \BibitemOpen
  \bibfield  {author} {\bibinfo {author} {\bibfnamefont {R.}~\bibnamefont {Loudon}},\ }\href@noop {} {\emph {\bibinfo {title} {The quantum theory of light}}}\ (\bibinfo  {publisher} {OUP Oxford},\ \bibinfo {year} {2000})\BibitemShut {NoStop}%
\bibitem [{\citenamefont {Wang}\ \emph {et~al.}(1997)\citenamefont {Wang}, \citenamefont {Gould},\ and\ \citenamefont {Stwalley}}]{K_natural}%
  \BibitemOpen
  \bibfield  {author} {\bibinfo {author} {\bibfnamefont {H.}~\bibnamefont {Wang}}, \bibinfo {author} {\bibfnamefont {P.~L.}\ \bibnamefont {Gould}},\ and\ \bibinfo {author} {\bibfnamefont {W.~C.}\ \bibnamefont {Stwalley}},\ }\bibfield  {title} {{\selectlanguage {english}\bibinfo {title} {Long-range interaction of the {39K} \textit{(4s)+} {39K} \textit{(4p)} asymptote by photoassociative spectroscopy. {I}. {The} g-pure long-range state and the long-range potential constants}},\ }\href {https://doi.org/10.1063/1.473804} {\bibfield  {journal} {\bibinfo  {journal} {The Journal of Chemical Physics}\ }\textbf {\bibinfo {volume} {106}},\ \bibinfo {pages} {7899} (\bibinfo {year} {1997})}\BibitemShut {NoStop}%
\bibitem [{\citenamefont {Fridman}(2012)}]{Saha}%
  \BibitemOpen
  \bibfield  {author} {\bibinfo {author} {\bibfnamefont {A.~A.}\ \bibnamefont {Fridman}},\ }\href@noop {} {\emph {\bibinfo {title} {Plasma chemistry}}},\ \bibinfo {edition} {first paperback edition}\ ed.\ (\bibinfo  {publisher} {Cambridge University Press},\ \bibinfo {address} {Cambridge},\ \bibinfo {year} {2012})\BibitemShut {NoStop}%
\bibitem [{\citenamefont {Saloman}\ and\ \citenamefont {Sansonetti}(2004)}]{Ne}%
  \BibitemOpen
  \bibfield  {author} {\bibinfo {author} {\bibfnamefont {E.~B.}\ \bibnamefont {Saloman}}\ and\ \bibinfo {author} {\bibfnamefont {C.~J.}\ \bibnamefont {Sansonetti}},\ }\bibfield  {title} {{\selectlanguage {english}\bibinfo {title} {Wavelengths, {Energy} {Level} {Classifications}, and {Energy} {Levels} for the {Spectrum} of {Neutral} {Neon}}},\ }\href {https://doi.org/10.1063/1.1797771} {\bibfield  {journal} {\bibinfo  {journal} {Journal of Physical and Chemical Reference Data}\ }\textbf {\bibinfo {volume} {33}},\ \bibinfo {pages} {1113} (\bibinfo {year} {2004})}\BibitemShut {NoStop}%
\bibitem [{\citenamefont {Sansonetti}\ and\ \citenamefont {Martin}(2005)}]{Ne_ionization}%
  \BibitemOpen
  \bibfield  {author} {\bibinfo {author} {\bibfnamefont {J.~E.}\ \bibnamefont {Sansonetti}}\ and\ \bibinfo {author} {\bibfnamefont {W.~C.}\ \bibnamefont {Martin}},\ }\bibfield  {title} {{\selectlanguage {english}\bibinfo {title} {Handbook of {Basic} {Atomic} {Spectroscopic} {Data}}},\ }\href {https://doi.org/10.1063/1.1800011} {\bibfield  {journal} {\bibinfo  {journal} {Journal of Physical and Chemical Reference Data}\ }\textbf {\bibinfo {volume} {34}},\ \bibinfo {pages} {1559} (\bibinfo {year} {2005})}\BibitemShut {NoStop}%
\bibitem [{\citenamefont {Olivero}\ and\ \citenamefont {Longbothum}(1977)}]{Olivero}%
  \BibitemOpen
  \bibfield  {author} {\bibinfo {author} {\bibfnamefont {J.}~\bibnamefont {Olivero}}\ and\ \bibinfo {author} {\bibfnamefont {R.}~\bibnamefont {Longbothum}},\ }\bibfield  {title} {{\selectlanguage {english}\bibinfo {title} {Empirical fits to the {Voigt} line width: {A} brief review}},\ }\href {https://doi.org/10.1016/0022-4073(77)90161-3} {\bibfield  {journal} {\bibinfo  {journal} {Journal of Quantitative Spectroscopy and Radiative Transfer}\ }\textbf {\bibinfo {volume} {17}},\ \bibinfo {pages} {233} (\bibinfo {year} {1977})}\BibitemShut {NoStop}%
\bibitem [{\citenamefont {Amici}\ \emph {et~al.}(2022)\citenamefont {Amici}, \citenamefont {Spiller}, \citenamefont {Ansalone},\ and\ \citenamefont {Miller}}]{wildfire_temp1}%
  \BibitemOpen
  \bibfield  {author} {\bibinfo {author} {\bibfnamefont {S.}~\bibnamefont {Amici}}, \bibinfo {author} {\bibfnamefont {D.}~\bibnamefont {Spiller}}, \bibinfo {author} {\bibfnamefont {L.}~\bibnamefont {Ansalone}},\ and\ \bibinfo {author} {\bibfnamefont {L.}~\bibnamefont {Miller}},\ }\bibfield  {title} {{\selectlanguage {english}\bibinfo {title} {Wildfires {Temperature} {Estimation} by {Complementary} {Use} of {Hyperspectral} {PRISMA} and {Thermal} ({ECOSTRESS} \& {L8})}},\ }\href {https://doi.org/10.1029/2022JG007055} {\bibfield  {journal} {\bibinfo  {journal} {Journal of Geophysical Research: Biogeosciences}\ }\textbf {\bibinfo {volume} {127}},\ \bibinfo {pages} {e2022JG007055} (\bibinfo {year} {2022})}\BibitemShut {NoStop}%
\bibitem [{\citenamefont {Creely}\ \emph {et~al.}(2020)\citenamefont {Creely}, \citenamefont {Greenwald}, \citenamefont {Ballinger}, \citenamefont {Brunner}, \citenamefont {Canik}, \citenamefont {Doody}, \citenamefont {Fülöp}, \citenamefont {Garnier}, \citenamefont {Granetz}, \citenamefont {Gray},\ and\ \citenamefont {et~al.}}]{Overview_Tokamak}%
  \BibitemOpen
  \bibfield  {author} {\bibinfo {author} {\bibfnamefont {A.~J.}\ \bibnamefont {Creely}}, \bibinfo {author} {\bibfnamefont {M.~J.}\ \bibnamefont {Greenwald}}, \bibinfo {author} {\bibfnamefont {S.~B.}\ \bibnamefont {Ballinger}}, \bibinfo {author} {\bibfnamefont {D.}~\bibnamefont {Brunner}}, \bibinfo {author} {\bibfnamefont {J.}~\bibnamefont {Canik}}, \bibinfo {author} {\bibfnamefont {J.}~\bibnamefont {Doody}}, \bibinfo {author} {\bibfnamefont {T.}~\bibnamefont {Fülöp}}, \bibinfo {author} {\bibfnamefont {D.~T.}\ \bibnamefont {Garnier}}, \bibinfo {author} {\bibfnamefont {R.}~\bibnamefont {Granetz}}, \bibinfo {author} {\bibfnamefont {T.~K.}\ \bibnamefont {Gray}},\ and\ \bibinfo {author} {\bibnamefont {et~al.}},\ }\bibfield  {title} {\bibinfo {title} {Overview of the sparc tokamak},\ }\href {https://doi.org/10.1017/S0022377820001257} {\bibfield  {journal} {\bibinfo  {journal} {Journal of Plasma Physics}\ }\textbf {\bibinfo {volume} {86}},\ \bibinfo {pages} {865860502} (\bibinfo {year} {2020})}\BibitemShut
  {NoStop}%
\bibitem [{\citenamefont {Greenwald}\ \emph {et~al.}(1988)\citenamefont {Greenwald}, \citenamefont {Terry}, \citenamefont {Wolfe}, \citenamefont {Ejima}, \citenamefont {Bell}, \citenamefont {Kaye},\ and\ \citenamefont {Neilson}}]{Greenwald_density}%
  \BibitemOpen
  \bibfield  {author} {\bibinfo {author} {\bibfnamefont {M.}~\bibnamefont {Greenwald}}, \bibinfo {author} {\bibfnamefont {J.}~\bibnamefont {Terry}}, \bibinfo {author} {\bibfnamefont {S.}~\bibnamefont {Wolfe}}, \bibinfo {author} {\bibfnamefont {S.}~\bibnamefont {Ejima}}, \bibinfo {author} {\bibfnamefont {M.}~\bibnamefont {Bell}}, \bibinfo {author} {\bibfnamefont {S.}~\bibnamefont {Kaye}},\ and\ \bibinfo {author} {\bibfnamefont {G.}~\bibnamefont {Neilson}},\ }\bibfield  {title} {\bibinfo {title} {A new look at density limits in tokamaks},\ }\href {https://doi.org/10.1088/0029-5515/28/12/009} {\bibfield  {journal} {\bibinfo  {journal} {Nuclear Fusion}\ }\textbf {\bibinfo {volume} {28}},\ \bibinfo {pages} {2199} (\bibinfo {year} {1988})}\BibitemShut {NoStop}%
\bibitem [{\citenamefont {Griem}(1997)}]{Plasma_spectroscopy}%
  \BibitemOpen
  \bibfield  {author} {\bibinfo {author} {\bibfnamefont {H.~R.}\ \bibnamefont {Griem}},\ }\href@noop {} {\emph {\bibinfo {title} {Principles of Plasma Spectroscopy}}},\ Cambridge Monographs on Plasma Physics\ (\bibinfo  {publisher} {Cambridge University Press},\ \bibinfo {year} {1997})\BibitemShut {NoStop}%
\bibitem [{\citenamefont {Kado}\ \emph {et~al.}(2011)\citenamefont {Kado}, \citenamefont {Suzuki}, \citenamefont {Iida},\ and\ \citenamefont {Muraki}}]{Stark_Tokyo}%
  \BibitemOpen
  \bibfield  {author} {\bibinfo {author} {\bibfnamefont {S.}~\bibnamefont {Kado}}, \bibinfo {author} {\bibfnamefont {K.}~\bibnamefont {Suzuki}}, \bibinfo {author} {\bibfnamefont {Y.}~\bibnamefont {Iida}},\ and\ \bibinfo {author} {\bibfnamefont {A.}~\bibnamefont {Muraki}},\ }\bibfield  {title} {{\selectlanguage {english}\bibinfo {title} {Doppler and {Stark} broadenings of spectral lines of highly excited helium atoms for measurement of detached recombining plasmas in {MAP}-{II} divertor simulator}},\ }\href {https://doi.org/10.1016/j.jnucmat.2011.01.048} {\bibfield  {journal} {\bibinfo  {journal} {Journal of Nuclear Materials}\ }\textbf {\bibinfo {volume} {415}},\ \bibinfo {pages} {S1174} (\bibinfo {year} {2011})}\BibitemShut {NoStop}%
\bibitem [{\citenamefont {Griem}\ and\ \citenamefont {Barr}(1975)}]{Stark_Griem}%
  \BibitemOpen
  \bibfield  {author} {\bibinfo {author} {\bibfnamefont {H.~R.}\ \bibnamefont {Griem}}\ and\ \bibinfo {author} {\bibfnamefont {W.~L.}\ \bibnamefont {Barr}},\ }\bibfield  {title} {\bibinfo {title} {Spectral line broadening by plasmas},\ }\href {https://api.semanticscholar.org/CorpusID:30091930} {\bibfield  {journal} {\bibinfo  {journal} {IEEE Transactions on Plasma Science}\ }\textbf {\bibinfo {volume} {3}},\ \bibinfo {pages} {227} (\bibinfo {year} {1975})}\BibitemShut {NoStop}%
\bibitem [{\citenamefont {Neyskens}\ \emph {et~al.}(2015)\citenamefont {Neyskens}, \citenamefont {Van~Eck}, \citenamefont {Jorissen}, \citenamefont {Goriely}, \citenamefont {Siess},\ and\ \citenamefont {Plez}}]{Star_temp2}%
  \BibitemOpen
  \bibfield  {author} {\bibinfo {author} {\bibfnamefont {P.}~\bibnamefont {Neyskens}}, \bibinfo {author} {\bibfnamefont {S.}~\bibnamefont {Van~Eck}}, \bibinfo {author} {\bibfnamefont {A.}~\bibnamefont {Jorissen}}, \bibinfo {author} {\bibfnamefont {S.}~\bibnamefont {Goriely}}, \bibinfo {author} {\bibfnamefont {L.}~\bibnamefont {Siess}},\ and\ \bibinfo {author} {\bibfnamefont {B.}~\bibnamefont {Plez}},\ }\bibfield  {title} {{\selectlanguage {english}\bibinfo {title} {The temperature and chronology of heavy-element synthesis in low-mass stars}},\ }\href {https://doi.org/10.1038/nature14050} {\bibfield  {journal} {\bibinfo  {journal} {Nature}\ }\textbf {\bibinfo {volume} {517}},\ \bibinfo {pages} {174} (\bibinfo {year} {2015})}\BibitemShut {NoStop}%
\bibitem [{\citenamefont {Burdonov}\ \emph {et~al.}(2020)\citenamefont {Burdonov}, \citenamefont {Revet}, \citenamefont {Bonito}, \citenamefont {Argiroffi}, \citenamefont {Béard}, \citenamefont {Bolanõs}, \citenamefont {Cerchez}, \citenamefont {Chen}, \citenamefont {Ciardi}, \citenamefont {Espinosa}, \citenamefont {Filippov}, \citenamefont {Pikuz}, \citenamefont {Rodriguez}, \citenamefont {Šmíd}, \citenamefont {Starodubtsev}, \citenamefont {Willi}, \citenamefont {Orlando},\ and\ \citenamefont {Fuchs}}]{star_density1}%
  \BibitemOpen
  \bibfield  {author} {\bibinfo {author} {\bibfnamefont {K.}~\bibnamefont {Burdonov}}, \bibinfo {author} {\bibfnamefont {G.}~\bibnamefont {Revet}}, \bibinfo {author} {\bibfnamefont {R.}~\bibnamefont {Bonito}}, \bibinfo {author} {\bibfnamefont {C.}~\bibnamefont {Argiroffi}}, \bibinfo {author} {\bibfnamefont {J.}~\bibnamefont {Béard}}, \bibinfo {author} {\bibfnamefont {S.}~\bibnamefont {Bolanõs}}, \bibinfo {author} {\bibfnamefont {M.}~\bibnamefont {Cerchez}}, \bibinfo {author} {\bibfnamefont {S.~N.}\ \bibnamefont {Chen}}, \bibinfo {author} {\bibfnamefont {A.}~\bibnamefont {Ciardi}}, \bibinfo {author} {\bibfnamefont {G.}~\bibnamefont {Espinosa}}, \bibinfo {author} {\bibfnamefont {E.}~\bibnamefont {Filippov}}, \bibinfo {author} {\bibfnamefont {S.}~\bibnamefont {Pikuz}}, \bibinfo {author} {\bibfnamefont {R.}~\bibnamefont {Rodriguez}}, \bibinfo {author} {\bibfnamefont {M.}~\bibnamefont {Šmíd}}, \bibinfo {author} {\bibfnamefont {M.}~\bibnamefont {Starodubtsev}}, \bibinfo {author} {\bibfnamefont {O.}~\bibnamefont
  {Willi}}, \bibinfo {author} {\bibfnamefont {S.}~\bibnamefont {Orlando}},\ and\ \bibinfo {author} {\bibfnamefont {J.}~\bibnamefont {Fuchs}},\ }\bibfield  {title} {\bibinfo {title} {Laboratory evidence for an asymmetric accretion structure upon slanted matter impact in young stars},\ }\href {https://doi.org/10.1051/0004-6361/202038189} {\bibfield  {journal} {\bibinfo  {journal} {Astronomy \& Astrophysics}\ }\textbf {\bibinfo {volume} {642}},\ \bibinfo {pages} {A38} (\bibinfo {year} {2020})}\BibitemShut {NoStop}%
\bibitem [{\citenamefont {Sahal-Bréchot}\ and\ \citenamefont {Elabidi}(2021)}]{star_density2}%
  \BibitemOpen
  \bibfield  {author} {\bibinfo {author} {\bibfnamefont {S.}~\bibnamefont {Sahal-Bréchot}}\ and\ \bibinfo {author} {\bibfnamefont {H.}~\bibnamefont {Elabidi}},\ }\bibfield  {title} {\bibinfo {title} {Stark broadening for {Br} {VI} and {Kr} {V}-{VII} lines in hot star atmospheres},\ }\href {https://doi.org/10.1051/0004-6361/202140729} {\bibfield  {journal} {\bibinfo  {journal} {Astronomy \& Astrophysics}\ }\textbf {\bibinfo {volume} {652}},\ \bibinfo {pages} {A47} (\bibinfo {year} {2021})}\BibitemShut {NoStop}%
\bibitem [{\citenamefont {Frémat}\ \emph {et~al.}(2023)\citenamefont {Frémat}, \citenamefont {Royer}, \citenamefont {Marchal}, \citenamefont {Blomme}, \citenamefont {Sartoretti}, \citenamefont {Guerrier}, \citenamefont {Panuzzo}, \citenamefont {Katz}, \citenamefont {Seabroke}, \citenamefont {Thévenin}, \citenamefont {Cropper}, \citenamefont {Benson}, \citenamefont {Damerdji}, \citenamefont {Haigron}, \citenamefont {Lobel}, \citenamefont {Smith}, \citenamefont {Baker}, \citenamefont {Chemin}, \citenamefont {David}, \citenamefont {Dolding}, \citenamefont {Gosset}, \citenamefont {Janßen}, \citenamefont {Jasniewicz}, \citenamefont {Plum}, \citenamefont {Samaras}, \citenamefont {Snaith}, \citenamefont {Soubiran}, \citenamefont {Vanel}, \citenamefont {Zorec}, \citenamefont {Zwitter}, \citenamefont {Brouillet}, \citenamefont {Caffau}, \citenamefont {Crifo}, \citenamefont {Fabre}, \citenamefont {Fragkoudi}, \citenamefont {Huckle}, \citenamefont {Lasne}, \citenamefont {Leclerc}, \citenamefont
  {Mastrobuono-Battisti}, \citenamefont {Jean-Antoine~Piccolo},\ and\ \citenamefont {Viala}}]{Rotational_linewidth}%
  \BibitemOpen
  \bibfield  {author} {\bibinfo {author} {\bibfnamefont {Y.}~\bibnamefont {Frémat}}, \bibinfo {author} {\bibfnamefont {F.}~\bibnamefont {Royer}}, \bibinfo {author} {\bibfnamefont {O.}~\bibnamefont {Marchal}}, \bibinfo {author} {\bibfnamefont {R.}~\bibnamefont {Blomme}}, \bibinfo {author} {\bibfnamefont {P.}~\bibnamefont {Sartoretti}}, \bibinfo {author} {\bibfnamefont {A.}~\bibnamefont {Guerrier}}, \bibinfo {author} {\bibfnamefont {P.}~\bibnamefont {Panuzzo}}, \bibinfo {author} {\bibfnamefont {D.}~\bibnamefont {Katz}}, \bibinfo {author} {\bibfnamefont {G.~M.}\ \bibnamefont {Seabroke}}, \bibinfo {author} {\bibfnamefont {F.}~\bibnamefont {Thévenin}}, \bibinfo {author} {\bibfnamefont {M.}~\bibnamefont {Cropper}}, \bibinfo {author} {\bibfnamefont {K.}~\bibnamefont {Benson}}, \bibinfo {author} {\bibfnamefont {Y.}~\bibnamefont {Damerdji}}, \bibinfo {author} {\bibfnamefont {R.}~\bibnamefont {Haigron}}, \bibinfo {author} {\bibfnamefont {A.}~\bibnamefont {Lobel}}, \bibinfo {author} {\bibfnamefont {M.}~\bibnamefont
  {Smith}}, \bibinfo {author} {\bibfnamefont {S.~G.}\ \bibnamefont {Baker}}, \bibinfo {author} {\bibfnamefont {L.}~\bibnamefont {Chemin}}, \bibinfo {author} {\bibfnamefont {M.}~\bibnamefont {David}}, \bibinfo {author} {\bibfnamefont {C.}~\bibnamefont {Dolding}}, \bibinfo {author} {\bibfnamefont {E.}~\bibnamefont {Gosset}}, \bibinfo {author} {\bibfnamefont {K.}~\bibnamefont {Janßen}}, \bibinfo {author} {\bibfnamefont {G.}~\bibnamefont {Jasniewicz}}, \bibinfo {author} {\bibfnamefont {G.}~\bibnamefont {Plum}}, \bibinfo {author} {\bibfnamefont {N.}~\bibnamefont {Samaras}}, \bibinfo {author} {\bibfnamefont {O.}~\bibnamefont {Snaith}}, \bibinfo {author} {\bibfnamefont {C.}~\bibnamefont {Soubiran}}, \bibinfo {author} {\bibfnamefont {O.}~\bibnamefont {Vanel}}, \bibinfo {author} {\bibfnamefont {J.}~\bibnamefont {Zorec}}, \bibinfo {author} {\bibfnamefont {T.}~\bibnamefont {Zwitter}}, \bibinfo {author} {\bibfnamefont {N.}~\bibnamefont {Brouillet}}, \bibinfo {author} {\bibfnamefont {E.}~\bibnamefont {Caffau}}, \bibinfo
  {author} {\bibfnamefont {F.}~\bibnamefont {Crifo}}, \bibinfo {author} {\bibfnamefont {C.}~\bibnamefont {Fabre}}, \bibinfo {author} {\bibfnamefont {F.}~\bibnamefont {Fragkoudi}}, \bibinfo {author} {\bibfnamefont {H.~E.}\ \bibnamefont {Huckle}}, \bibinfo {author} {\bibfnamefont {Y.}~\bibnamefont {Lasne}}, \bibinfo {author} {\bibfnamefont {N.}~\bibnamefont {Leclerc}}, \bibinfo {author} {\bibfnamefont {A.}~\bibnamefont {Mastrobuono-Battisti}}, \bibinfo {author} {\bibfnamefont {A.}~\bibnamefont {Jean-Antoine~Piccolo}},\ and\ \bibinfo {author} {\bibfnamefont {Y.}~\bibnamefont {Viala}},\ }\bibfield  {title} {\bibinfo {title} {\textit{{Gaia}} {Data} {Release} 3: {Properties} of the line-broadening parameter derived with the {Radial} {Velocity} {Spectrometer} ({RVS})},\ }\href {https://doi.org/10.1051/0004-6361/202243809} {\bibfield  {journal} {\bibinfo  {journal} {Astronomy \& Astrophysics}\ }\textbf {\bibinfo {volume} {674}},\ \bibinfo {pages} {A8} (\bibinfo {year} {2023})}\BibitemShut {NoStop}%
\bibitem [{\citenamefont {Padovani}\ \emph {et~al.}(2017)\citenamefont {Padovani}, \citenamefont {Alexander}, \citenamefont {Assef}, \citenamefont {De~Marco}, \citenamefont {Giommi}, \citenamefont {Hickox}, \citenamefont {Richards}, \citenamefont {Smolčić}, \citenamefont {Hatziminaoglou}, \citenamefont {Mainieri},\ and\ \citenamefont {Salvato}}]{AGN_bands}%
  \BibitemOpen
  \bibfield  {author} {\bibinfo {author} {\bibfnamefont {P.}~\bibnamefont {Padovani}}, \bibinfo {author} {\bibfnamefont {D.~M.}\ \bibnamefont {Alexander}}, \bibinfo {author} {\bibfnamefont {R.~J.}\ \bibnamefont {Assef}}, \bibinfo {author} {\bibfnamefont {B.}~\bibnamefont {De~Marco}}, \bibinfo {author} {\bibfnamefont {P.}~\bibnamefont {Giommi}}, \bibinfo {author} {\bibfnamefont {R.~C.}\ \bibnamefont {Hickox}}, \bibinfo {author} {\bibfnamefont {G.~T.}\ \bibnamefont {Richards}}, \bibinfo {author} {\bibfnamefont {V.}~\bibnamefont {Smolčić}}, \bibinfo {author} {\bibfnamefont {E.}~\bibnamefont {Hatziminaoglou}}, \bibinfo {author} {\bibfnamefont {V.}~\bibnamefont {Mainieri}},\ and\ \bibinfo {author} {\bibfnamefont {M.}~\bibnamefont {Salvato}},\ }\bibfield  {title} {{\selectlanguage {english}\bibinfo {title} {Active galactic nuclei: what’s in a name?}},\ }\href {https://doi.org/10.1007/s00159-017-0102-9} {\bibfield  {journal} {\bibinfo  {journal} {The Astronomy and Astrophysics Review}\ }\textbf {\bibinfo
  {volume} {25}},\ \bibinfo {pages} {2} (\bibinfo {year} {2017})}\BibitemShut {NoStop}%
\bibitem [{\citenamefont {Akylas}\ and\ \citenamefont {Georgantopoulos}(2021)}]{AGN_temp1}%
  \BibitemOpen
  \bibfield  {author} {\bibinfo {author} {\bibfnamefont {A.}~\bibnamefont {Akylas}}\ and\ \bibinfo {author} {\bibfnamefont {I.}~\bibnamefont {Georgantopoulos}},\ }\bibfield  {title} {\bibinfo {title} {Distribution of the coronal temperature in {Seyfert} 1 galaxies},\ }\href {https://doi.org/10.1051/0004-6361/202141186} {\bibfield  {journal} {\bibinfo  {journal} {Astronomy \& Astrophysics}\ }\textbf {\bibinfo {volume} {655}},\ \bibinfo {pages} {A60} (\bibinfo {year} {2021})}\BibitemShut {NoStop}%
\bibitem [{\citenamefont {Foschini}(2002)}]{AGN_turbo}%
  \BibitemOpen
  \bibfield  {author} {\bibinfo {author} {\bibfnamefont {L.}~\bibnamefont {Foschini}},\ }\bibfield  {title} {\bibinfo {title} {On the broadening of emission lines in active galactic nuclei},\ }\href {https://doi.org/10.1051/0004-6361:20020231} {\bibfield  {journal} {\bibinfo  {journal} {Astronomy \& Astrophysics}\ }\textbf {\bibinfo {volume} {385}},\ \bibinfo {pages} {62} (\bibinfo {year} {2002})}\BibitemShut {NoStop}%
\bibitem [{\citenamefont {{Kollatschny, W.}}\ and\ \citenamefont {{Zetzl, M.}}(2013)}]{AGN_width}%
  \BibitemOpen
  \bibfield  {author} {\bibinfo {author} {\bibnamefont {{Kollatschny, W.}}}\ and\ \bibinfo {author} {\bibnamefont {{Zetzl, M.}}},\ }\bibfield  {title} {\bibinfo {title} {The shape of broad-line profiles in active galactic nuclei},\ }\href {https://doi.org/10.1051/0004-6361/201219411} {\bibfield  {journal} {\bibinfo  {journal} {Astronomy\&Astrophysics}\ }\textbf {\bibinfo {volume} {549}},\ \bibinfo {pages} {A100} (\bibinfo {year} {2013})}\BibitemShut {NoStop}%
\bibitem [{\citenamefont {Liu}\ \emph {et~al.}(2020)\citenamefont {Liu}, \citenamefont {Ying}, \citenamefont {Zhong}, \citenamefont {Xu}, \citenamefont {Han}, \citenamefont {Yu},\ and\ \citenamefont {Cai}}]{heater_Lithium}%
  \BibitemOpen
  \bibfield  {author} {\bibinfo {author} {\bibfnamefont {X.}~\bibnamefont {Liu}}, \bibinfo {author} {\bibfnamefont {P.}~\bibnamefont {Ying}}, \bibinfo {author} {\bibfnamefont {X.}~\bibnamefont {Zhong}}, \bibinfo {author} {\bibfnamefont {J.}~\bibnamefont {Xu}}, \bibinfo {author} {\bibfnamefont {Y.}~\bibnamefont {Han}}, \bibinfo {author} {\bibfnamefont {S.}~\bibnamefont {Yu}},\ and\ \bibinfo {author} {\bibfnamefont {X.}~\bibnamefont {Cai}},\ }\bibfield  {title} {\bibinfo {title} {Highly efficient thermo-optic tunable micro-ring resonator based on an lnoi platform},\ }\href {https://doi.org/10.1364/OL.410192} {\bibfield  {journal} {\bibinfo  {journal} {Opt. Lett.}\ }\textbf {\bibinfo {volume} {45}},\ \bibinfo {pages} {6318} (\bibinfo {year} {2020})}\BibitemShut {NoStop}%
\bibitem [{\citenamefont {Sayem}\ \emph {et~al.}(2020)\citenamefont {Sayem}, \citenamefont {Cheng}, \citenamefont {Wang},\ and\ \citenamefont {Tang}}]{SNSPD1}%
  \BibitemOpen
  \bibfield  {author} {\bibinfo {author} {\bibfnamefont {A.~A.}\ \bibnamefont {Sayem}}, \bibinfo {author} {\bibfnamefont {R.}~\bibnamefont {Cheng}}, \bibinfo {author} {\bibfnamefont {S.}~\bibnamefont {Wang}},\ and\ \bibinfo {author} {\bibfnamefont {H.~X.}\ \bibnamefont {Tang}},\ }\bibfield  {title} {{\selectlanguage {english}\bibinfo {title} {Lithium-niobate-on-insulator waveguide-integrated superconducting nanowire single-photon detectors}},\ }\href {https://doi.org/10.1063/1.5142852} {\bibfield  {journal} {\bibinfo  {journal} {Applied Physics Letters}\ }\textbf {\bibinfo {volume} {116}},\ \bibinfo {pages} {151102} (\bibinfo {year} {2020})}\BibitemShut {NoStop}%
\bibitem [{\citenamefont {Prencipe}\ \emph {et~al.}(2023)\citenamefont {Prencipe}, \citenamefont {Gyger}, \citenamefont {Baghban}, \citenamefont {Zichi}, \citenamefont {Zeuner}, \citenamefont {Lettner}, \citenamefont {Schweickert}, \citenamefont {Steinhauer}, \citenamefont {Elshaari}, \citenamefont {Gallo},\ and\ \citenamefont {Zwiller}}]{SNSPD3}%
  \BibitemOpen
  \bibfield  {author} {\bibinfo {author} {\bibfnamefont {A.}~\bibnamefont {Prencipe}}, \bibinfo {author} {\bibfnamefont {S.}~\bibnamefont {Gyger}}, \bibinfo {author} {\bibfnamefont {M.~A.}\ \bibnamefont {Baghban}}, \bibinfo {author} {\bibfnamefont {J.}~\bibnamefont {Zichi}}, \bibinfo {author} {\bibfnamefont {K.~D.}\ \bibnamefont {Zeuner}}, \bibinfo {author} {\bibfnamefont {T.}~\bibnamefont {Lettner}}, \bibinfo {author} {\bibfnamefont {L.}~\bibnamefont {Schweickert}}, \bibinfo {author} {\bibfnamefont {S.}~\bibnamefont {Steinhauer}}, \bibinfo {author} {\bibfnamefont {A.~W.}\ \bibnamefont {Elshaari}}, \bibinfo {author} {\bibfnamefont {K.}~\bibnamefont {Gallo}},\ and\ \bibinfo {author} {\bibfnamefont {V.}~\bibnamefont {Zwiller}},\ }\bibfield  {title} {{\selectlanguage {english}\bibinfo {title} {Wavelength-{Sensitive} {Superconducting} {Single}-{Photon} {Detectors} on {Thin} {Film} {Lithium} {Niobate} {Waveguides}}},\ }\href {https://doi.org/10.1021/acs.nanolett.3c02324} {\bibfield  {journal} {\bibinfo  {journal}
  {Nano Letters}\ }\textbf {\bibinfo {volume} {23}},\ \bibinfo {pages} {9748} (\bibinfo {year} {2023})}\BibitemShut {NoStop}%
\bibitem [{\citenamefont {Lu}\ \emph {et~al.}(2021)\citenamefont {Lu}, \citenamefont {Al~Sayem}, \citenamefont {Gong}, \citenamefont {Surya}, \citenamefont {Zou},\ and\ \citenamefont {Tang}}]{Hong_OPO}%
  \BibitemOpen
  \bibfield  {author} {\bibinfo {author} {\bibfnamefont {J.}~\bibnamefont {Lu}}, \bibinfo {author} {\bibfnamefont {A.}~\bibnamefont {Al~Sayem}}, \bibinfo {author} {\bibfnamefont {Z.}~\bibnamefont {Gong}}, \bibinfo {author} {\bibfnamefont {J.~B.}\ \bibnamefont {Surya}}, \bibinfo {author} {\bibfnamefont {C.-L.}\ \bibnamefont {Zou}},\ and\ \bibinfo {author} {\bibfnamefont {H.~X.}\ \bibnamefont {Tang}},\ }\bibfield  {title} {{\selectlanguage {english}\bibinfo {title} {Ultralow-threshold thin-film lithium niobate optical parametric oscillator}},\ }\href {https://doi.org/10.1364/OPTICA.418984} {\bibfield  {journal} {\bibinfo  {journal} {Optica}\ }\textbf {\bibinfo {volume} {8}},\ \bibinfo {pages} {539} (\bibinfo {year} {2021})}\BibitemShut {NoStop}%
\bibitem [{\citenamefont {Stone}\ \emph {et~al.}(2024)\citenamefont {Stone}, \citenamefont {Westly}, \citenamefont {Moille},\ and\ \citenamefont {Srinivasan}}]{OPO_heater}%
  \BibitemOpen
  \bibfield  {author} {\bibinfo {author} {\bibfnamefont {J.}~\bibnamefont {Stone}}, \bibinfo {author} {\bibfnamefont {D.}~\bibnamefont {Westly}}, \bibinfo {author} {\bibfnamefont {G.}~\bibnamefont {Moille}},\ and\ \bibinfo {author} {\bibfnamefont {K.}~\bibnamefont {Srinivasan}},\ }\bibfield  {title} {{\selectlanguage {english}\bibinfo {title} {On-chip {Kerr} parametric oscillation with integrated heating for enhanced frequency tuning and control}},\ }\href {https://doi.org/10.1364/OL.523704} {\bibfield  {journal} {\bibinfo  {journal} {Optics Letters}\ }\textbf {\bibinfo {volume} {49}},\ \bibinfo {pages} {3118} (\bibinfo {year} {2024})}\BibitemShut {NoStop}%
\bibitem [{\citenamefont {Almeida}\ and\ \citenamefont {Lipson}(2004)}]{bistability}%
  \BibitemOpen
  \bibfield  {author} {\bibinfo {author} {\bibfnamefont {V.~R.}\ \bibnamefont {Almeida}}\ and\ \bibinfo {author} {\bibfnamefont {M.}~\bibnamefont {Lipson}},\ }\bibfield  {title} {\bibinfo {title} {Optical bistability on a silicon chip},\ }\href {https://doi.org/10.1364/OL.29.002387} {\bibfield  {journal} {\bibinfo  {journal} {Opt. Lett.}\ }\textbf {\bibinfo {volume} {29}},\ \bibinfo {pages} {2387} (\bibinfo {year} {2004})}\BibitemShut {NoStop}%
\bibitem [{\citenamefont {Walker}\ \emph {et~al.}(2020)\citenamefont {Walker}, \citenamefont {De~Vries}, \citenamefont {Felici},\ and\ \citenamefont {Schuster}}]{plasma_heat_eqn}%
  \BibitemOpen
  \bibfield  {author} {\bibinfo {author} {\bibfnamefont {M.~L.}\ \bibnamefont {Walker}}, \bibinfo {author} {\bibfnamefont {P.}~\bibnamefont {De~Vries}}, \bibinfo {author} {\bibfnamefont {F.}~\bibnamefont {Felici}},\ and\ \bibinfo {author} {\bibfnamefont {E.}~\bibnamefont {Schuster}},\ }\bibfield  {title} {\bibinfo {title} {Introduction to {Tokamak} {Plasma} {Control}},\ }in\ \href {https://doi.org/10.23919/ACC45564.2020.9147561} {\emph {\bibinfo {booktitle} {2020 {American} {Control} {Conference} ({ACC})}}}\ (\bibinfo  {publisher} {IEEE},\ \bibinfo {address} {Denver, CO, USA},\ \bibinfo {year} {2020})\ pp.\ \bibinfo {pages} {2901--2918}\BibitemShut {NoStop}%
\bibitem [{\citenamefont {AlOmar}(2020)}]{chebyshev}%
  \BibitemOpen
  \bibfield  {author} {\bibinfo {author} {\bibfnamefont {A.~S.}\ \bibnamefont {AlOmar}},\ }\bibfield  {title} {{\selectlanguage {english}\bibinfo {title} {Line width at half maximum of the {Voigt} profile in terms of {Gaussian} and {Lorentzian} widths: {Normalization}, asymptotic expansion, and chebyshev approximation}},\ }\href {https://doi.org/10.1016/j.ijleo.2019.163919} {\bibfield  {journal} {\bibinfo  {journal} {Optik}\ }\textbf {\bibinfo {volume} {203}},\ \bibinfo {pages} {163919} (\bibinfo {year} {2020})}\BibitemShut {NoStop}%
\bibitem [{\citenamefont {Gardner}\ \emph {et~al.}(2006)\citenamefont {Gardner}, \citenamefont {Mather}, \citenamefont {Clampin}, \citenamefont {Doyon}, \citenamefont {Greenhouse}, \citenamefont {Hammel}, \citenamefont {Hutchings}, \citenamefont {Jakobsen}, \citenamefont {Lilly}, \citenamefont {Long} \emph {et~al.}}]{Webb_motivation}%
  \BibitemOpen
  \bibfield  {author} {\bibinfo {author} {\bibfnamefont {J.~P.}\ \bibnamefont {Gardner}}, \bibinfo {author} {\bibfnamefont {J.~C.}\ \bibnamefont {Mather}}, \bibinfo {author} {\bibfnamefont {M.}~\bibnamefont {Clampin}}, \bibinfo {author} {\bibfnamefont {R.}~\bibnamefont {Doyon}}, \bibinfo {author} {\bibfnamefont {M.~A.}\ \bibnamefont {Greenhouse}}, \bibinfo {author} {\bibfnamefont {H.~B.}\ \bibnamefont {Hammel}}, \bibinfo {author} {\bibfnamefont {J.~B.}\ \bibnamefont {Hutchings}}, \bibinfo {author} {\bibfnamefont {P.}~\bibnamefont {Jakobsen}}, \bibinfo {author} {\bibfnamefont {S.~J.}\ \bibnamefont {Lilly}}, \bibinfo {author} {\bibfnamefont {K.~S.}\ \bibnamefont {Long}}, \emph {et~al.},\ }\bibfield  {title} {\bibinfo {title} {The james webb space telescope},\ }\href@noop {} {\bibfield  {journal} {\bibinfo  {journal} {Space Science Reviews}\ }\textbf {\bibinfo {volume} {123}},\ \bibinfo {pages} {485} (\bibinfo {year} {2006})}\BibitemShut {NoStop}%
\bibitem [{\citenamefont {Boyd}(2008)}]{Nonlinear_Boyd}%
  \BibitemOpen
  \bibfield  {author} {\bibinfo {author} {\bibfnamefont {R.~W.}\ \bibnamefont {Boyd}},\ }\href@noop {} {\emph {\bibinfo {title} {Nonlinear optics}}},\ \bibinfo {edition} {3rd}\ ed.\ (\bibinfo  {publisher} {Academic Press},\ \bibinfo {address} {Amsterdam ; Boston},\ \bibinfo {year} {2008})\BibitemShut {NoStop}%
\end{thebibliography}%

\end{document}